\documentclass[reprint,aps,pre]{revtex4-1}
\usepackage{graphicx}

\begin{document}
\author{Rasmus A. X. Persson}
\affiliation{Department of Chemistry \& Molecular Biology, University of
Gothenburg, SE-412 96 Gothenburg, Sweden}
\email{rasmusp@chem.gu.se}
\title{Perturbation method to calculate the density of states}

\begin{abstract}
Monte Carlo switching moves (``perturbations'') are defined between two or more
classical Hamiltonians sharing a common ground-state energy. The ratio of the
density of states (DOS) of one system to that of another is related to the
ensemble averages of the microcanonical acceptance probabilities of switching
between these Hamiltonians, analogously to the case of Bennett's acceptance
ratio method for the canonical ensemble [C. H. Bennett, \emph{J. Comput.
Phys.}, \textbf{22}, 245 (1976)].  Thus, if the DOS of one of the systems is
known, one obtains those of the others and, hence, the partition functions. As
a simple test case, the vapor pressure of an anharmonic Einstein crystal is
computed, using the harmonic Einstein crystal as the reference system in one
dimension; an auxiliary calculation is also performed in three dimensions. As a
further example of the algorithm, the energy dependence of the ratio of the DOS
of the square-well and hard-sphere tetradecamers is determined, from which the
temperature dependence of the constant-volume heat capacity of the square-well
system is calculated and compared with canonical Metropolis Monte Carlo
estimates. For these cases and reference systems, the perturbation calculations
exhibit a higher degree of convergence per Monte Carlo cycle than Wang-Landau
(WL) sampling, although for the one-dimensional oscillator the WL sampling is
ultimately more efficient for long runs. Last, we calculate the vapor pressure
of liquid gold using an empirical Sutton-Chen many-body potential and the ideal
gas as the reference state. Although this proves the general applicability of
the method, by its inherent perturbation approach, the algorithm is suitable
for those particular cases where the properties of a related system are well
known.
\end{abstract}



\maketitle

\section{Introduction}
A complete description of an isolated system at energy $E$ is given by the
phase space volume,
\begin{equation}
\label{eq:phasevol}
\Omega(E) \equiv \frac {1} {N! h^{3 N}} \int \mathrm d \vec q \mathrm d \vec p
\Theta(E - H(\vec q, \vec p)),
\end{equation}
where $\vec q, \vec p$ are $3N$-dimensional vectors stating the positions and
momenta, respectively, of the $N$ particles, $\Theta(x)$ the Heaviside step
function, $h$ Planck's constant and $H(\vec q, \vec p)$ the Hamiltonian.
Through this quantity---or the closely related density of states (DOS)
$\omega(E) = \partial \Omega / \partial E$---the connection with the entropy of
classical thermodynamics is established as one of $S \propto \ln \Omega(E)$
(Hertz definition) or $S \propto \ln \omega(E)$ (Planck definition).
These two definitions are not mathematically equivalent and, as pointed out and
discussed by Dunkel and Hilbert (see Ref. \onlinecite{dunkel06} and references
cited therein), there is disagreement in the literature as to which one is
correct.  However, these two definitions become {\em numerically} the same for
large systems. Indeed, if the system is great enough in the number of its
degrees of freedom, fluctuations in its kinetic energy will be vanishingly
small, the potential energy distribution will be Boltzmannian and the system
can be said to be at equilibrium at constant temperature, which is a desirable
situation as it can be reproduced more readily in reality, for which the
systems studied are generally large in this sense. Unfortunately, the size of
systems that can be investigated by computer simulation may still be far from
adequately approaching this limit.

The most common solution to this problem of size is to couple the system, in
the mathematical sense, to an infinitely large heat reservoir at constant
temperature, thereby creating a formally infinite system.  This system is
governed by the canonical partition function (CPF), which can be expressed as
\begin{eqnarray}
\label{eq:part}
Q = e^{U_0 / k T} \int_{U_0}^{\infty} \mathrm d E \omega(E) e^{-E / k T},
\end{eqnarray}
where $U_0$ is the lowest possible energy, $k$ Boltzmann's constant, and $T$
the absolute temperature.  This, or a mathematically equivalent, route to the
CPF has been exploited in numerical methods such as the reference system
equilibration (RSE) method \cite{ming96}, the histogramming \cite{labastie90,
cheng92} and multihistogram \cite{poteau94, calvo95} methods, the histogram
reweighting method \cite{ferrenberg88, ferrenberg89}, the Wang-Landau (WL)
sampling \cite{wang01a, wang01b}, multicanonical methods \cite{berg91, lee93,
geyer95}, transition matrix methods \cite{wang99, heilmann05} or the nested
sampling (NS) algorithm \cite{skilling04, skilling06, partay10, do11}.

The CPF is directly related to the free energy by
\begin{equation}
A(T, V) = - k T \ln Q(T, V),
\end{equation}
and thus knowledge of it enables one to compute the temperature or volume, $V$,
dependence of any desired thermodynamic property. Because the integrand
$\omega(E) e^{-E / k T}$ is a sharply peaked function in $E$, it is numerically
an easier task to obtain the partition function at a specific temperature than
to obtain the complete DOS. If one is interested in free energies only at one
or a few specific temperatures, {\em especially} low ones, direct methods
\cite{zwanzig54, hansen69, henderson70, torrie74, torrie77} to the free energy,
{\em i. e.} the CPF at pre-defined temperature, will always be more efficient,
simply because they have a smaller region of integration about which to worry.
The DOS approach, on the other hand, is more powerful when a range of
temperatures is of interest, especially in systems or models where the CPF has
no volume dependence, {\em e. g.} lattice models. The width of the temperature
interval of interest implicitly defines the width of corresponding energy
interval $[E_-, E_+]$ that one needs to consider in a numerical search for
$\omega(E)$. However, for continuous potentials, $\omega(E)$ approaches the
known DOS of the ideal gas at high $E$, as the kinetic energy contributions
will dominate the potential energy ones. In these cases, $E_+$ can be defined
independently of any temperature.

In the WL method, originally developed for model lattice systems but
generalized to continuous Hamiltonians by later authors \cite{yan02, shell02,
mauro07, desgranges09}, the DOS is computed through a random walk subject to
importance sampling whose weights are iteratively adjusted in an attempt to
make all energies equiprobable. The weights that achieve this are reciprocal to
the DOS. The precision by which the weights are adjusted is iteratively
increased until desired precision is reached. The main drawback of the method
is that the statistics to estimate the convergence of the weight update factors
needs to be gathered anew between each update of the precision, leading in some
sense to a ``duplication of efforts.'' A great wealth of literature has sprung
up around the WL method, and its extensions \cite{schulz02, dayal04,
troster05, lee06, poulain06, belardinelli07b, zhou08, cunha-netto08, soudan11,
dickman11}, and so it is widely known and recognized.  Therefore, we shall make
use of it for comparison purposes with the perturbation calculations.

In the NS method, efficacy is achieved by having the random walk subject to a
weight function that acts on a non-uniform distribution of energy segments,
concentrating on the low energy regions. The limits of each energy segment are
calculated ``on the fly,'' starting from high energies and proceeding downwards
by cutting the lower segments in two. The limits are set from the condition
that the probability of encountering a configuration belonging to a given
segment be equal to a predetermined function of the depth of that segment in
the partitioning tree. Once a limit is found, it stays fixed and is never
subject to reevaluation. This removes the ``duplication of efforts'' of the WL
algorithm. The most problematic case for this algorithm is for potentials that
exhibit large regions of infinite energy, as in for instance the square-well
fluid, because then the energy partitioning scheme cannot be gradual. There is
hence no benefit in using the simplification of the hard molecular core with
this method.  

In the RSE method, on the other hand, the system is coupled to a finite heat
bath with which it is allowed to exchange energy. The DOS of the heat bath
is presumed known. The combined system is evolved according to the
microcanonical probability distribution and the probability of the system of
having different energies $E < E_\mathrm{tot}$ is histogrammatically tracked
and related, up to an $E_\mathrm{tot}$-dependent factor, to the sought system
DOS. The main drawback is that the factor can only be calculated precisely for
very low energy and smooth potentials, and good statistics is only obtained in
a narrow range of $E$ below $E_\mathrm{tot}$, meaning in practice that several
simulations at different $E_\mathrm{tot}$ have to be run. By careful
considerations of the continuity of the DOS, the $E_\mathrm{tot}$-dependent
factor may be extrapolated to higher energies in the end.

In this Paper, we shall investigate an alternative method: a route to
obtaining $\omega(E)$ assuming, like the RSE method, that the DOS of a
different system is known. Unlike the RSE method, however, the idea is that the
other system is also similar, and thus knowledge of its DOS is able to speed up
the calculation by focusing on the difference between the two systems. This
lessens the need for importance sampling; in essence, the one system is used to
sample the important regions of the other system, because these regions are
supposedly shared to a large extent because of systemic similarity. We shall
develop the method in the next Section, and after that examine some simple
numerical examples of its use.

\section{Description of the algorithm}
Consider two systems, for simplicity labeled as $0$ and $1$, for which the
DOS are $\omega_0(E)$ (known) and $\omega_1(E)$ (sought). Classical
microcanonical sampling of either of these systems can be carried out
efficiently if the potential energy depends on the configuration only. Under
this restriction, one simulates a Markov process in configuration space using
the acceptance probability \cite{severin78, schranz91, ray91},
\begin{equation}
\label{eq:acc}
P_E(U_i, U_f) = \min \left \{ \frac {(E - U_f)^{M N / 2 - 1}} {(E - U_i)^{M N
/ 2 - 1}}, 1 \right \}
\end{equation}
if $E > U_f$ and zero otherwise, where $U_i$ and $U_f$ denote the potential
energies before and after, respectively, an unbiased trial move in
configuration space. This equation represents the ratio between the densities
of the kinetic energy states for a $M$-dimensional configuration space with
$N$ molecules, and is the proper weight function to use for the
microcanonical ensemble where all accessible states are considered
equiprobable.

Let us now suppose that systems $0$ and $1$ are ``similar'' in the sense that
they share the same configuration space. Then system $1$, differing only by its
potential energy expression, can be viewed as the result of a perturbation on
system $0$. For instance, let us define the generalized Hamiltonian,
\begin{equation}
\label{eq:ham}
H_\lambda(\vec q, \vec p) = H_0(\vec q, \vec p) + \lambda U'(\vec q),
\end{equation}
where $U'(\vec q)$ is the perturbation depending on $\vec q$ only, and
$\lambda$ an interpolating factor between the reference ($\lambda = 0$) and
fully perturbed ($\lambda = 1$) system. Correspondingly, we may define
$\Omega_\lambda(E)$ according to eq.~(\ref{eq:phasevol}) after inserting
eq.~(\ref{eq:ham}). Let us now consider the superensemble that includes
$\lambda$ as an extra coordinate. Tentatively, we write its phase space volume
as,
\begin{equation}
\widehat \Omega(\widehat E) \propto \int \mathrm d \vec q \mathrm d \vec p
\mathrm d \lambda \mathrm d \zeta \Theta \left (\widehat E - H_0(\vec q, \vec p)
- \lambda U'(\vec q) - \zeta^2 / 2 \eta \right ),
\end{equation}
where $\eta$ is a formal mass associated with the $\lambda$-motion and
$\zeta$ is a dummy variable of integration for the formal momentum of this
motion. The total energy of this ensemble is $\widehat E$ which is different
from the regular total energy $E$ because it also contains a kinetic
contribution $\eta \dot{\lambda}^2 / 2$ in addition to that of the regular
coordinates. Therefore, we consider instead the \emph{constrained}
superensemble, whose phase space volume is
\begin{eqnarray}
\widehat \Omega'(\widehat E) & \propto & \int \mathrm d \vec q \mathrm d \vec p
\mathrm d \lambda \Theta \left (\widehat E - H_0(\vec q, \vec p) - \lambda
U'(\vec q) \right ) \nonumber \\
& = & \int \mathrm d \lambda \Omega_\lambda(E),
\end{eqnarray}
for which $\zeta \equiv 0$, and so $\widehat E = E$ as required. We then take
the derivative with respect to $E$ and obtain the constrained DOS of the
superensemble, 
\begin{equation}
\widehat \omega'(E) \propto \int \mathrm d \lambda \omega_\lambda(E).
\end{equation}
Because the weighting factor of this integral is $\lambda$-independent, the
acceptance probability of each state is still only proportional to the density
of its kinetic energy states, and we see that the random walk also along the
$\lambda$-coordinate should be governed by an unmodified eq.~(\ref{eq:acc}).

We now define an ``equilibrium constant'' $K_{ij}(E)$ as the ratio between the
number of cycles the Markov chain sampling the constrained superensemble visits
system $\lambda = \lambda_j$ to the number of cycles it visits system $\lambda
= \lambda_i$. The existence of this equilibrium constant is guaranteed by the
detailed balance condition that the Markov chain fulfills. From the direct
proportionality between the microcanonical probability and the DOS it follows
that
\begin{equation}
\label{eq:equconst}
K_{ij}(E) = \frac {\omega_{\lambda_j}(E)}
{\omega_{\lambda_i}(E)} = \frac {\langle P_{ij}(E) \rangle_i} {\langle
P_{ji}(E) \rangle_j},
\end{equation}
where $P_{ij}$ is the acceptance probability for changing from $\lambda =
\lambda_i$ to $\lambda = \lambda_j$, and $\langle \ldots \rangle_i$ denotes a
microcanonical ensemble average over the system $\lambda = \lambda_i$. The last
equality follows from the flux balance at equilibrium: to wit that the gross
flux between two states, given by the product of the acceptance probability and
the occupation probability, is equal but of opposing sign in the reverse
directions, \textit{i.~e.},
\begin{equation}
\label{eq:balance}
P_{E}(U_i, U_j) (E - U_i)^{M N / 2 - 1} = P_{E}(U_j, U_i) (E - U_j)^{M N / 2 -
1},
\end{equation}
where $U_i$ and $U_j$ denote the potential energies of two states defined by
any two arbitrary sets of the non-momentum coordinates of the Hamiltonian; in
the argument to follow, we restrict our attention to when $i$ and $j$
correspond to $\lambda_i$ and $\lambda_j$, irrespective of $\vec q$. The
quantity $(E - U)^{M N / 2 - 1}$ is proportional to the kinetic energy DOS, and
because the configuration space is sampled subject to this bias, it is thus
also proportional to the probability of being in a state of potential energy
$U$. The probability of being in any spatial configuration with $\lambda =
\lambda_i$ is proportional to $\omega_i(E)$. Let us therefore divide
eq.~(\ref{eq:balance}) by $\omega_i(E) \omega_j(E)$ and integrate over the
spatial dimensions,
\begin{eqnarray}
\int \mathrm d \vec q \frac {P_{E}(U_i(\vec q), U_j(\vec q)) (E - U_i(\vec
q))^{M N / 2 - 1}}
{\omega_i(E) \omega_j(E)} & = & \nonumber \\ 
\int \mathrm d \vec q \frac {P_{E}(U_j(\vec q), U_i(\vec q)) (E - U_j(\vec
q))^{M N / 2 - 1}} {\omega_i(E) \omega_j(E)}.
\end{eqnarray}
We now identify the microcanonical ensemble average of $P_E$ on each side as,
\begin{equation}
\langle P_{ij}(E) \rangle_i  = \int \mathrm d \vec q \frac {P_{E}(U_i(\vec
q), U_j(\vec q)) (E - U_i(\vec q))^{M N / 2 - 1}} {\omega_i(E)},
\end{equation}
and then obtain eq.~(\ref{eq:equconst}) after rearrangement. It is at this
point appropriate to stress that in the case when $M N / 2 = 1$,
eq.~(\ref{eq:equconst}) does not hold and the algorithm, as here outlined, is
not applicable. Such is the case of a single particle ($N = 1$)
confined to two spatial dimensions ($M = 2$); it is never the case in three
spatial dimensions ($M = 3$). Excepting that special case, we have that
\begin{equation}
\label{eq:eqlbrm}
K_{01}(E) = \prod_{i=0}^{i_\mathrm{max} - 1} K_{i,i+1} = \frac {\langle
P_{01}(E) \rangle_0} {\langle P_{10}(E) \rangle_1} = \frac {\omega_1(E)}
{\omega_0(E)},
\end{equation}
and it is from this relation that $\omega_1(E)$ may be extracted, if
$\omega_0(E)$ is known, in addition to the ratio $\omega_1(E) / \omega_0(E)$
which is always obtained. The method outlined may be regarded as a special
case of Bennett's method \cite{bennett76}, but applied to the microcanonical
ensemble. Alternatively, if one does not sample the transition probabilities,
but instead propagates the system between the two states, it may also be
regarded as a case of the expanded ensemble method \cite{lyubartsev92} applied
to the microcanonical ensemble; this, however, is a line of attack which we
shall not pursue.

Implicit in the derivation so far is that the minimum value of the potential
energy expression is to be independent of $\lambda$. In other words, the
``potential energy'' of a configuration $(\vec q, \lambda)$ is to be understood
as the potential energy {\em difference} with respect to the global potential
minimum over $\vec q$ keeping $\lambda$ constant. This follows from
eq.~(\ref{eq:balance}) (unless we make $P_E$ explicitly $\lambda$-dependent)
and the following argument. In eq.~(\ref{eq:balance}), the quantity $E - U$ is
the kinetic energy. Let us say that the maximum kinetic energy is $E$, obtained
when $U = U_0$, the minimum potential energy. As the $\lambda$-coordinate does
not affect the kinetic energy, the potential energy $U = U_0$ should correspond
to the maximum kinetic energy $E$ regardless of the value of $\lambda$. This
does not restrict the method in any formal sense but it may pose a practical
hurdle for very complicated Hamiltonians for which energy minimization is
difficult. This is especially true if several intermediate $\lambda$-values are
considered over a chain of gradual perturbations, if these affect the energy
minimum in a non-trivial way.

\subsection{Schematic of the algorithm}
In the simplest case, one considers only two systems: reference (system 0) and
perturbed system (system $\lambda$, where $\lambda$ indicates the degree of
perturbation). Given a starting configuration $\{\vec q_i\}$ in
the phase space of system 0 with the potential energy $U_i$, the algorithm may
be outlined as follows when broken down into its elementary steps. 
\begin{enumerate}
\item Generate uniform random number $a \in [0, 1]$.
\item If $a \geq B$, go to step \ref{item:step}.
\item Generate random configuration $\{\vec q_f\}$ by random displacement from
$\{\vec q_i\}$.
\item Calculate the energy $U_f$ of $\{\vec q_f\}$.
\item Generate uniform random number $a' \in [0, 1]$.
\item If $a' \leq P_E(U_i, U_f)$ by eq.~[\ref{eq:acc}], let $\{\vec q_i\} \to
\{\vec q_f\}$.
\item Iterate from step 1.
\item {\label{item:step} Calculate the energy $U_\lambda$ of perturbed system
in configuration $\{\vec q_i\}$.}
\item Calculate $P_E(U_i, U_\lambda)$ and accumulate average $\langle
P_{0\lambda} \rangle_0 \equiv \langle P_E(U_i, U_\lambda) \rangle_0$.
\item Are averages converged? If not, iterate from step 1.
\end{enumerate}
Here $B \in [0, 1]$ is an arbitrary constant that decides the priority among
the computer cycles for either propagating the microcanonical system, to insure
that the average acceptance probabilities sampled come from more or less
uncorrelated points, or sampling the actual averages, to insure they get
sufficient statistics. In general, the same steps have to be carried out also
for system $\lambda$ on system 0, but in some specific cases (as we shall see
below) this is not necessary, because $P_{\lambda 0}(E) \equiv 1$ in these
cases. When one runs the algorithm on many systems, as when one considers
gradual perturbations, or for many discrete $E$-values, it is natural to run
them in parallel for maximum efficacy. 

The convergence and accuracy of the proposed method hinge on the accuracy in
the ensemble averages $\langle P_{\lambda 0} \rangle_\lambda$ and $\langle P_{0
\lambda} \rangle_0$. Since obviously both $\lim_{\lambda \to 0} P_{0
\lambda}(E) \equiv 1$ and $\lim_{\lambda \to 0} P_{\lambda 0}(E) \equiv 1$ hold
for all $E$ and are thus without statistical uncertainty in this limit, it
is certain to state that for any system sufficiently close to the reference
system, the perturbation calculations will always be superior to direct
methods. In this Paper, for the most part convergence has been deemed to have
been achieved when $\langle P_{\lambda 0} \rangle_\lambda / \langle
P_{0\lambda} \rangle_0$ differs by less than an amount $\epsilon > 0$ from its
previous value calculated a fixed number of cycles earlier. A more stringent
alternative that leaves less room for apparent convergence by chance, and thus
a more efficient simulation, would be to require that the estimated standard
error of each individual average is below some threshold, but the simple
convergence criterion has proven satisfactory in the cases considered.  

\section{Numerical examples}
In principle, any two Hamiltonians for which we can calculate the
requisite ensemble averages $\langle P_{01} \rangle_0$ and $\langle P_{10}
\rangle_1$ can be used in this method. The algorithm hence does not pose any
greater programming challenges than that of regular Metropolis Monte Carlo
techniques. Here we shall consider two primary cases: the square-well fluid and
a class of anharmonic Einstein crystals; and one secondary example: the vapor
pressure of liquid gold. The simplicity of the Einstein crystal is motivated by
a desire to keep the computational demands low as repeated comparisons with
other methods quickly become prohibitively expensive otherwise; the
generalization to more degrees of freedom is trivial in all other respects.
The square-well fluid, on the other hand, presents an interesting test case in
that appreciable regions of its configuration space are of infinite potential
energy. Last but not least, the simplicity of the primary numerical examples
makes it easier to gauge the correct behavior to be exhibited by the
calculation. Nevertheless, the example calculation of the vapor pressure of
gold illustrates the general applicability of the method.

Except for the calculations on gold (see Section~\ref{sec:gold}), all of the
numerical examples to be presented have been compiled using the GNU C Compiler
(version 4.4.3) with its intrinsic random number generator and executed on a
single 2 GHz core of the author's ``Intel Core 2 Duo'' laptop computer. Memory
demands of the calculations are all insignificant.

\subsection{Anharmonic Einstein crystal}
We will investigate in this section some simple numerical test cases on a class
of anharmonic Einstein crystals. By this we mean three-dimensional crystals for
which the CPF of $N$ molecules can be written $Q(T) = q(T)^{3 N / M}$, where
$q(T)$ is the partition function of a single $M$-dimensional oscillator with $M
= 1, 2, 3$ being the spatial (not the phase space) dimensionality. The
anharmonic systems studied were governed by the potential energy expression
\begin{equation}
\label{eq:osc}
u_\lambda(x) = x^2 + \lambda x^4,
\end{equation}
where the term $\lambda x^4$ is the perturbation and $x$ the displacement from
equilibrium. 

\begin{figure}
\includegraphics{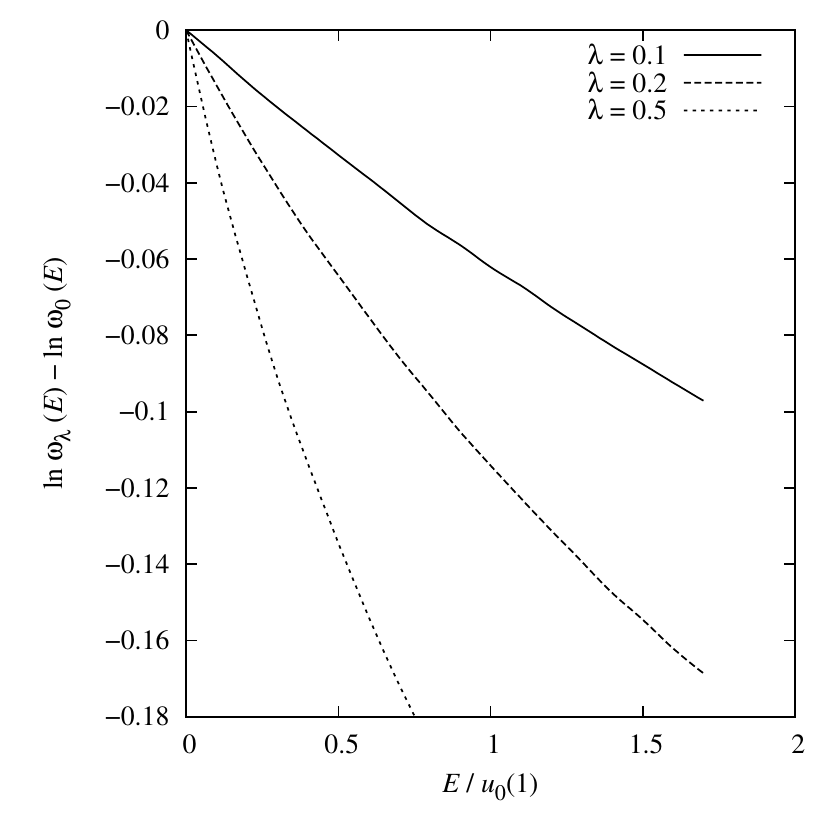}
\caption{Logarithm of the DOS of oscillators subject to the potential energy of
eq. (\ref{eq:osc}) for $\lambda = 0.1, 0.2, 0.5$ given with respect to the
$\lambda = 0.0$ reference DOS in one spatial dimension.}
\label{fig:dos}
\end{figure}

\begin{figure}
\includegraphics{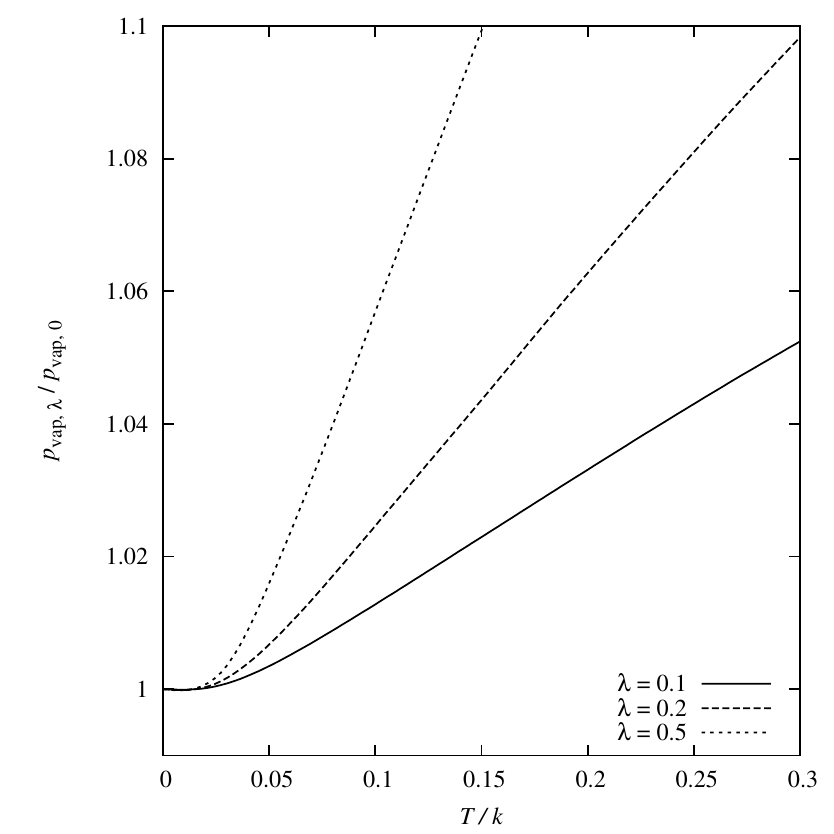}
\caption{Deviation of the vapor pressure of the anharmonic Einstein crystal
from the harmonic reference.}
\label{fig:vap}
\end{figure}

Let us first consider the case when $M = 1$, because it is the simplest. The
natural reference system to use when approaching the anharmonic Einstein
crystal is that of the harmonic Einstein crystal since $q(T)$ for an harmonic
oscillator is known analytically: its (classical) DOS, 
\begin{equation}
\omega_0(E) = \frac {2 \pi} {h} \sqrt{\frac {m} {k_f}},
\end{equation}
is independent of energy. In this equation, $m$ is the mass and $k_f$ the force
constant. Visual inspection (Figure \ref{fig:dos}) reveals that the logarithms
of the resulting DOS are quasi-linear in energy with a constant of
proportionality directly proportional to $\lambda$. The vapor is assumed ideal,
so that the vapor pressure is given by (assuming one-dimensional $q$),
\begin{equation}
\label{eq:einstein}
p_\mathrm{vap}(T) = \frac {k T} {(\Lambda q)^3},
\end{equation}
where $\Lambda$ is the thermal de Broglie wavelength. This equation is derived
from the equality between the chemical potentials of the gas and crystal phases
in the limit of $N \to \infty$. The results of these calculations are shown in
Figure \ref{fig:vap} plotted against temperature.  They are useful as a
``yardstick'' of how large the perturbations considered here are in relation to
real systems.

\begin{figure}
\includegraphics{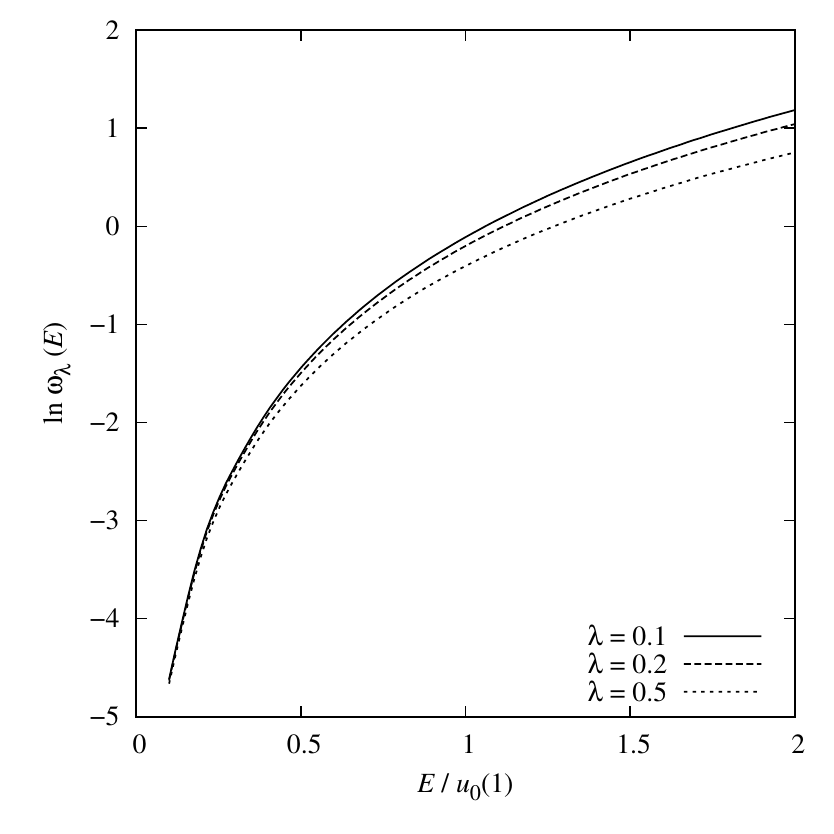}
\caption{Logarithm of the DOS of oscillators subject to the potential energy of
eq. (\ref{eq:osc}) for $\lambda = 0.1, 0.2, 0.5$ (arbitrary units) in three
spatial dimensions.}
\label{fig:dos2}
\end{figure}

Let us now briefly consider the case when $M = 3$. The DOS of the reference
system is then $E$-dependent,
\begin{equation}
\label{eq:dos3d}
\omega_0(E) = 4 \left (\frac {\pi} {h} \sqrt{\frac {m} {k_f}} \right)^3 E^2.
\end{equation}
The calculated DOS as a function of energy are shown in Figure~\ref{fig:dos2}.
The most striking thing about this calculation is the much quicker convergence
times for $M = 3$ than for $M = 1$. The reason is that only one ensemble
average has to be sampled in this case, namely $\langle P_{01}(E) \rangle_0$,
$\langle P_{10}(E) \rangle_1$ being unity for all $E$ and this holds also for
$M > 3$. Not only does this bring about a two-fold speed increase because of
the reduced workload, but also the statistical uncertainty in the ratio between
the two averages is reduced: both because only one average now has statistical
uncertainty and also because the individual terms of this average are all
non-zero, whereas in the one-dimensional case some terms in the averaging were
strictly zero, leading to very oscillatory terms. The net effect is a quicker
convergence and will be quantitatively assessed below.

\subsubsection{Comparison of efficacy}
We now turn to a comparison with the WL algorithm in terms of accuracy and
speed of convergence. In the discussion to follow, the DOS has been restricted
to 17 energy grid points, spaced $0.1u_0(1)$ units apart. The maximum random
displacement in $x$ was $1.5$ for both the WL sampling and perturbation
calculations for $M = 1$ and $1.0$ for $M = 3$. For the WL sampling, also the
random walk in momentum space used this maximum displacement with the mass
taken to be unity. To combat boundary artifacts in the WL simulations, the
update rule of Schulz \textit{et al.} \cite{schulz03} was employed. The
perturbation calculations do not suffer from any boundary artifacts.

\begin{figure}
\includegraphics{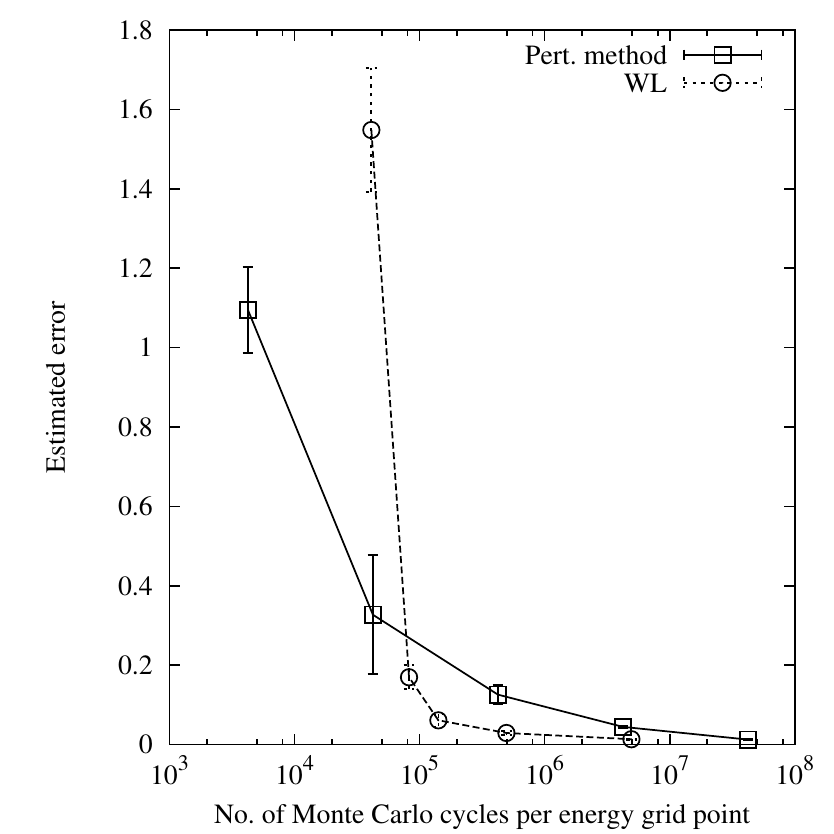}
\caption{Estimated errors ($\langle |\frac {\mathrm d \ln
\omega_\mathrm{est.}} {\mathrm d E} - \frac {\mathrm d \ln
\omega_\mathrm{exact}} {\mathrm d E}| \rangle$) as a function
of number of Monte Carlo cycles (normalized per number of energy grid points)
from runs of the one-dimensional anharmonic oscillator with $\lambda =
0.1$ and 17 energy grid points. In the graph, the number of Monte Carlo cycles
reported for the perturbation calculations is the actual number times two to
correct for that in each cycle two ensemble averages are sampled. Error bars
represent standard deviations from three independent runs.}
\label{fig:comp3}
\end{figure}

\begin{figure}
\includegraphics{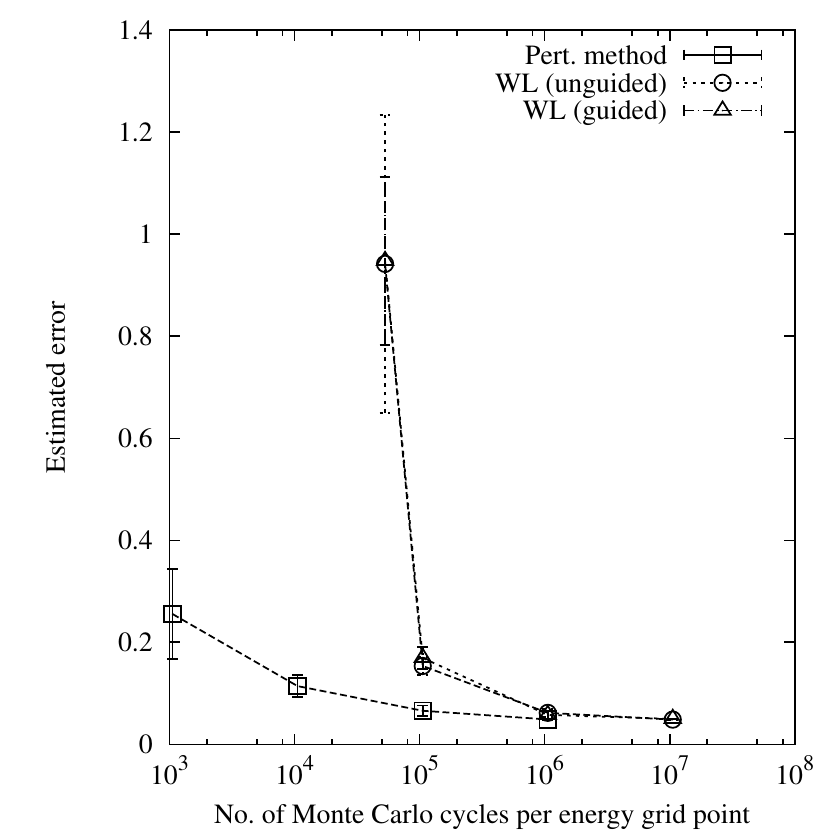}
\caption{Estimated errors ($\langle |\frac {\mathrm d \ln
\omega_\mathrm{est.}} {\mathrm d E} - \frac {\mathrm d \ln
\omega_\mathrm{exact}} {\mathrm d E}| \rangle$) as a function
of the number of Monte Carlo cycles (normalized per number of energy grid
points) for the three-dimensional anharmonic oscillator with
$\lambda = 0.1$ and 17 energy grid points. The qualifiers of ``guided'' and
``unguided'', respectively, denote whether the WL sampling was initialized from
the reciprocal DOS of the reference state or a ``blank slate''. The apparent
limiting error reflects the disagreement in the second decimal place about what
the limiting average of $\frac {\mathrm d \ln \omega} {\mathrm d E}$ is
according to the two methods. See text for details.}
\label{fig:comp}
\end{figure}

\begin{figure}
\includegraphics{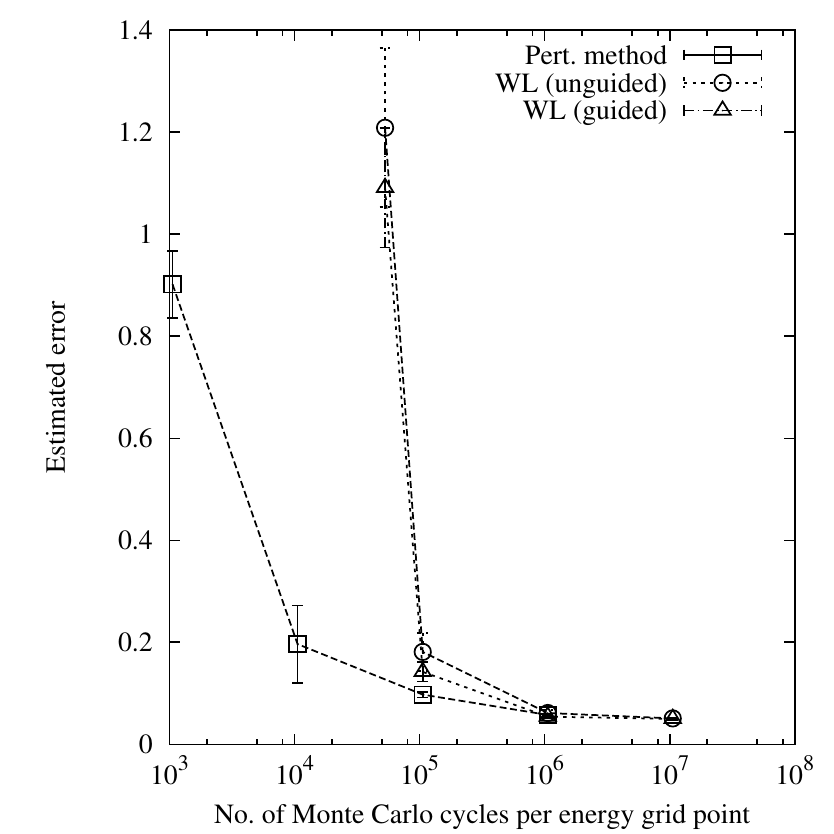}
\caption{Same as Figure~\ref{fig:comp} but for $\lambda = 0.5$.}
\label{fig:comp2}
\end{figure}

As an objective measure of the convergence, we calculate the mean deviation of
the numerical energy derivative $\frac {\mathrm d \ln \omega(E)} {\mathrm d E}$
from that of the ``exact'' DOS. When calculating the error of the WL sampling,
the ``exact'' DOS used for reference has been calculated by long runs of the
perturbation method; likewise, when calculating the error of the perturbation
method, the comparison is made with respect to the DOS that has been
calculated by long runs of the WL sampling.  Running shorter simulations with
either the perturbation or WL method allows these partially converged results
to be compared against the reference curve produced by the other method and
used as an indication of the level of convergence attained. The results of
these comparisons are given in Figures~\ref{fig:comp3}, \ref{fig:comp} and
\ref{fig:comp2}.  The absolute values of $\ln \omega_\lambda(E)$ cannot be
compared directly as the WL sampling does only provide $\omega_\lambda(E)$ up
to an undetermined multiplicative constant that is unique to each run. 

As a check, the results of these two methods have been found to agree roughly
up to the second decimal place in $\frac {\mathrm d \ln \omega_\lambda} {\mathrm
d E}$ on average for $M = 3$ but for $M = 1$, a higher
degree of agreement between the two methods is not a problem. To analyze the
source of this discrepancy---which on the face of it would seem to indicate
that at least one of the algorithms exhibits convergence difficulties to the
exact result---we note that whereas the perturbation calculation yields
$\omega(E)$ at discrete energy grid points $\{E_i\}$, the WL sampling rather
calculates the average $\langle \omega(E) \rangle$ over intervals $\Delta E =
0.1 u_0(1)$ centered on each $E_i$. The numerical, finite-difference
derivatives do hence not agree between the two methods, unless $\omega(E)$
happens to be a linear or nearly linear function. In any case, this level of
accuracy is sufficient for our purposes as it clearly allows us to judge which
method is faster.  

For the perturbation method, there is in principle the question of what the
optimal distribution of labor is between sampling the averages required to
calculate the ratio of the DOS, and how often one propagates the microcanonical
Markov chains. No claim is made that the division, of (80\% probability)
propagating the microcanonical Markov chain and (20\% probability) sampling
averages, employed in this comparison is optimal. It is outside the scope of
this work to provide a detailed analysis of this optimum, or of the influence
of step size, acceptance rates and so on. We note that similar issues are also
present for the WL algorithm as, indeed, the rate and reliability of
convergence of the WL algorithm has been the subject of much discussion in the
literature.  In fact, the measure of convergence as originally proposed may
lead to convergence difficulties \cite{yan03, belardinelli07a, komura12}.
Indeed, for the case $M = 1$, when run using the requirement that the histogram
should be ``flat'', the WL algorithm is noticeably slower and does not achieve
a smooth curve as reliably as the perturbation method (data not shown).  

However, if one instead employs the convergence criterion for the WL sampling
that was suggested by Morozov and Lin \cite{morozov07, morozov09} where, rather
than enforce strict ``flatness'' of the sampled histogram, we require a minimum
number of ``visits'' for each histogram entry before updating the WL precision
factor, the WL sampling---still for $M = 1$---is \emph{quicker} than the
perturbation method. The convergence according to this criterion was tested
every $10^5$:th cycle and the required minimum number of visits was
\begin{equation}
H_i = \frac {\ln 2} {2 \ln f_i}
\end{equation}
for the $i$:th iteration of the WL sampling. In this equation $f_i$ is the
multiplicative precision factor used by the WL algorithm in adjusting the
estimate of the DOS. In this implementation, it is given by $f_i =
\sqrt{f_{i-1}}$, with $f_0 = e$.  In this comparison the WL algorithm was
initiated from the flat DOS of the harmonic oscillator, and so benefits to the
exact same extent as the perturbation method from the similarity between the
two systems.  

Let us now comment on the case when $M = 3$. In this case, we find the opposite
results as compared to the one-dimensional case in terms of the rate of
convergence, namely that the perturbation method is quicker than the WL
algorithm, even in the long run. For a perfectly fair comparison, the
convergence of the WL sampling has also been investigated when the algorithm is
initialized from the reciprocal of eq.~(\ref{eq:dos3d}), instead of from a
``blank slate''. As can be seen in Figures~\ref{fig:comp} and~\ref{fig:comp2},
the effect is small, within the error bars and mainly confined to the early
cycles. This means that initializing the WL sampling from the DOS of the
reference system is not a viable alternative to the actual perturbation method.

\subsection{Square-well fluid tetradecamer} 
\label{sec:sw}
In this numerical example, we consider the square-well fluid as a perturbation
of its hard-sphere analog. The DOS of the hard-sphere gas obeys the form
$\omega(E) = \xi(N, V) E^{3 N / 2 - 1}$, where $\xi(N, V)$ is an unknown
function of the number of particles and volume. Our ignorance of the precise
form of this constant of proportionality means that only the ratio between the
DOS will be possible to provide completely. This, in turn, means that we cannot
compute, for instance, the phase diagram of the square-well fluid, but only
properties at constant $N$ and $V$. A prime example of such a property is the
constant-volume heat capacity. The unit of energy and temperature that we will
use for the remainder of this section is the magnitude of the pair potential at
unit distance and $\lambda = 1$. The unit of length is the hard-core diameter.

To be precise, the reference system interacts through the pair potential
\begin{equation}
u_0(r) = \left \{ \begin{array}{l r} \infty & r < 1 \\ 0 & r
\geq 1 \end{array} \right .,
\end{equation}
where $r$ denotes the intermolecular separation. The perturbation we introduce
is
\begin{equation}
u'(r) = \left \{ \begin{array}{l r} -1 & r < \sigma \\ 0 & r \geq \sigma
\end{array} \right .,
\end{equation}
so that the total pair interaction is written 
\begin{equation}
u_\lambda(r) = u_0(r) + \lambda u'(r).
\end{equation}
For our chosen combination of systems, with the common energy zero-level,
transitions from the perturbed system to the reference are always accepted,
which means that $\langle P_{10}(E) \rangle_1 \equiv 1$ and so there is, unlike
in the previous section, no need to consider two ensembles explicitly. Thus,
only configurations of the reference hard-sphere system have to be generated,
and furthermore, these configurations are independent of $E$ so that all
averages $\langle P_{01}(E) \rangle_0$ can be sampled simultaneously for a
given density. In the calculations to follow, a random 95\% of the Monte Carlo
cycles consisted of propagating the microcanonical Markov chain and the
remaining 5\% of accumulating averages.

We let $\sigma = \sqrt{2}$, an arbitrary choice based purely on aesthetic
appeal: it is the lattice constant of the close-packed cubic crystal. For this
system, the energy minimum is $-52 \lambda$ for 14 molecules. Thus, we consider
the total energy expression,
\begin{equation}
U_\lambda(\{r_{ij}\}) = 52 \lambda + \sum_{i>j=1}^{14} u_\lambda(r_{ij}),
\end{equation}
whose zero-level is independent of $\lambda$.  In the preceding equation,
$\{r_{ij}\}$ is the ordered set of all pairwise distances between the fourteen
molecules. To satisfy the requirements of the microcanonical ensemble, we
introduce the constraint that the cluster is confined to a fixed spherical
volume, arbitrarily chosen to be either $500 \pi / 3$, {\em i. e.}
corresponding to a radius of $5$, and a volume fraction of $1.4\%$ (``low
density''); or $108 \pi / 3$, corresponding to a radius of $3$, and a volume
fraction of $7 / 108$ (``low-medium density''); or $9 \pi / 2$, corresponding
to a radius of $3/2$ and a volume fraction of $14 / 27$ (``high density'').

For the propagation of the hard-core Markov chain, one molecule was moved at a
time. At the volume fraction of $1.4\%$, the displacement step was
$3.0$; at the volume fraction of $7 / 108$, it was $1.0$; and at the volume
fraction of $14 / 27$, it was $0.15$. These displacements led to acceptance
rates of 51\%, 54\% and 52\%, respectively. The DOS was sampled in energy
intervals of $1.4$, starting at $E = 32$ for the low-medium density and
covering the
higher energies in batches of 60 grid points. The calculations proceeded for at
least $2 \times 10^8$ cycles, which on the author's machine took a little less
than 3 minutes of real time for all 60 energy points sampled at once on a
single processor core, but considerably longer runs were found necessary to
achieve the same level of high convergence in the lowest energy regions, where
up to 20 minutes could be necessary. A refined attack would distribute the
energy grid unequally over the energy range.

\begin{figure}
\includegraphics{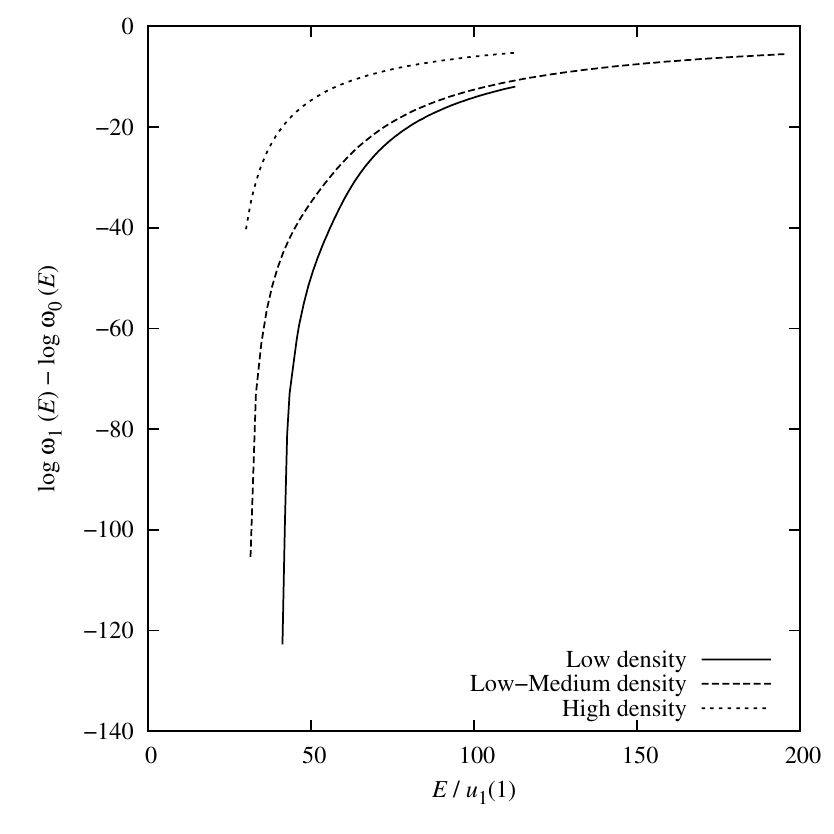}
\caption{Difference between the logarithms of the DOS of the square-well and
hard-sphere tetradecamers at three different densities.}
\label{fig:swdos}
\end{figure}

\begin{figure}
\includegraphics{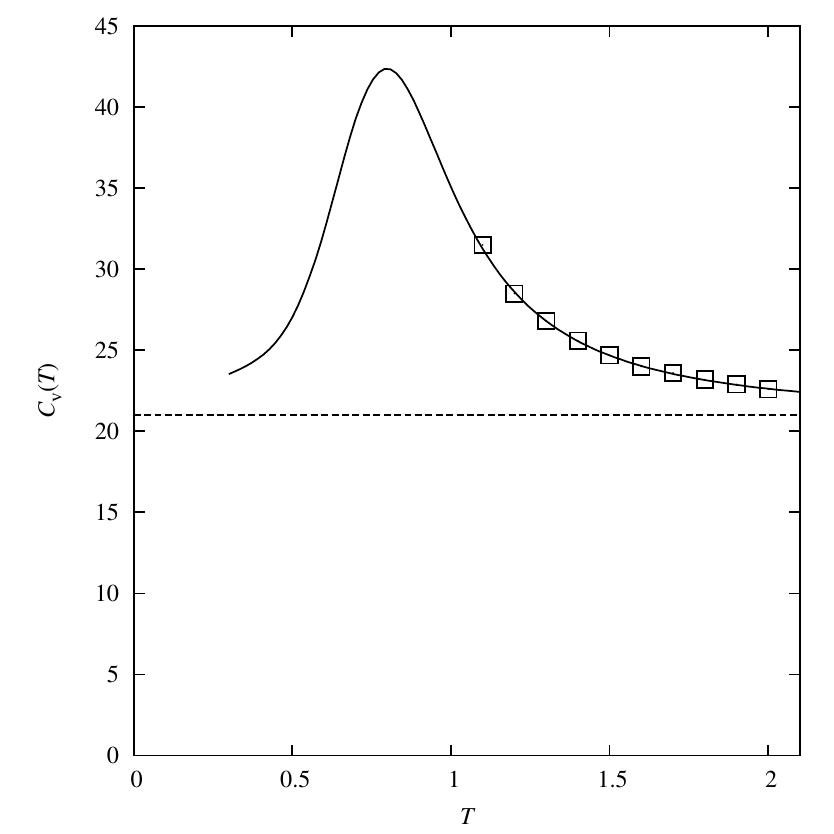}
\caption{Constant-volume heat capacity of the square-well tetradecamer as a
function of temperature at a volume fraction of 7 / 108. The broad peak in the
heat capacity is indicative of a first-order phase transition of a finite-sized
system. The dashed line is the translational equipartition heat capacity of 21,
as well as the heat capacity of the hard-sphere reference system. Squares
denote heat capacities calculated from canonical Monte Carlo simulations
according to eq.~(\ref{eq:fluct}).}
\label{fig:swdos2}
\end{figure}

One interesting aspect of the way we have defined the perturbation in $\lambda$
is the self-similarity that arises. Consider $\omega_a(E)$, where $a$ is any
point along the $\lambda$-axis. This quantity is
given by,
\begin{equation}
\omega_a(E) = \omega_0(E) \langle P_{0 a}(E) \rangle_0,
\end{equation}
because $\langle P_{a 0}(E) \rangle_\lambda \equiv 1$ in this system. But,
\begin{equation}
\langle P_{0 a}(E) \rangle_0 = \left \langle \frac {E / a - U} {E /
a} \right \rangle_0 = \langle P_{01}(E / a) \rangle_0.
\end{equation}
Therefore, we have the self-similarity relation,\footnote{This scaling formula
holds for most potentials when the perturbation is defined like this.}
\begin{equation}
\label{eq:selfsimilar}
\frac {\omega_{a}(E)} {\omega_0(E)} = \frac {\omega_1(E / a)}
{\omega_0(E / a)}.
\end{equation}
We see through this formula when we take the limit $a \to 0$ that
$\omega_1(E) \to \omega_0(E)$, when $E \to \infty$. An indication that the
computer code is well and working is that the DOS for the square-well
tetradecamer (given on the logarithmic scale with respect to the reference
system in Figure~\ref{fig:swdos}) actually shows this mathematically proven
convergence on that of the hard-sphere tetradecamer at high energies.  The
algorithm also runs quicker until convergence in those cases. The interesting
part, where convergence is also a bit more problematic, is for the low energy
regions where the DOS of the square-well tetradecamer exhibits a clear
deviation from its hard-sphere counterpart. The depth of this ``dip'' in the
curve is decreased when the density is increased. It is easy to see why this
should be by considering the close-packed density where the molecules have no
liberty of movement left, conditions under which the hard-sphere and the
square-well fluid are indistinguishable.

In Figure~\ref{fig:swdos2} is shown the temperature dependence of the
constant-volume heat capacity of the coupled system at the ``low-medium''
density corresponding to the volume fraction of $7 / 108$. The heat capacity
was calculated through the statistical mechanical relation
\begin{equation}
C_\mathrm v = 2 k T \frac {\partial \ln Q} {\partial T} + k T^2 \frac
{\partial^2 \ln Q} {\partial T^2}.
\end{equation}
The broad peak in this function at around $T \approx 0.9$ is characteristic of
a first-order phase transition far from the thermodynamic limit
\cite{westergren03}, in contradistinction with the singularity that one obtains
for the infinite system, even if contrary to the case of Ref.
\onlinecite{westergren03} it is clear from the density in this case that it is
question of a gas-liquid rather than a liquid-solid transition. At high
temperatures, we expect the translational equipartition value of
$C_\mathrm v = 3N/2 = 21$ to hold, and this is borne out by the graph.
Moreover, this is also the limiting heat capacity at low temperatures, since
the law of Dulong and Petit does not hold for the square-well fluid. This is
because the potential is not analytical, and so there is no first-order
quadratic potential energy term to contribute to the heat capacity. This gives
rise to a largely symmetric peak in the heat capacity. For comparison, the heat
capacity calculated from regular constant-volume Monte Carlo simulations and
the fluctuation formula,
\begin{equation}
\label{eq:fluct}
C_\mathrm v = \frac {3 N k} {2} + \frac {\langle U^2 \rangle - \langle U
\rangle^2} {k T^2},
\end{equation}
are also shown in Figure~\ref{fig:swdos2}. It is to be noted that these
simulations are very difficult to converge in the low-temperature regime, not
the least because of the numerical instability that arises from the $T^2$
denominator for small $T$.

\subsubsection{Comparison of efficacy}
\begin{figure}
\includegraphics{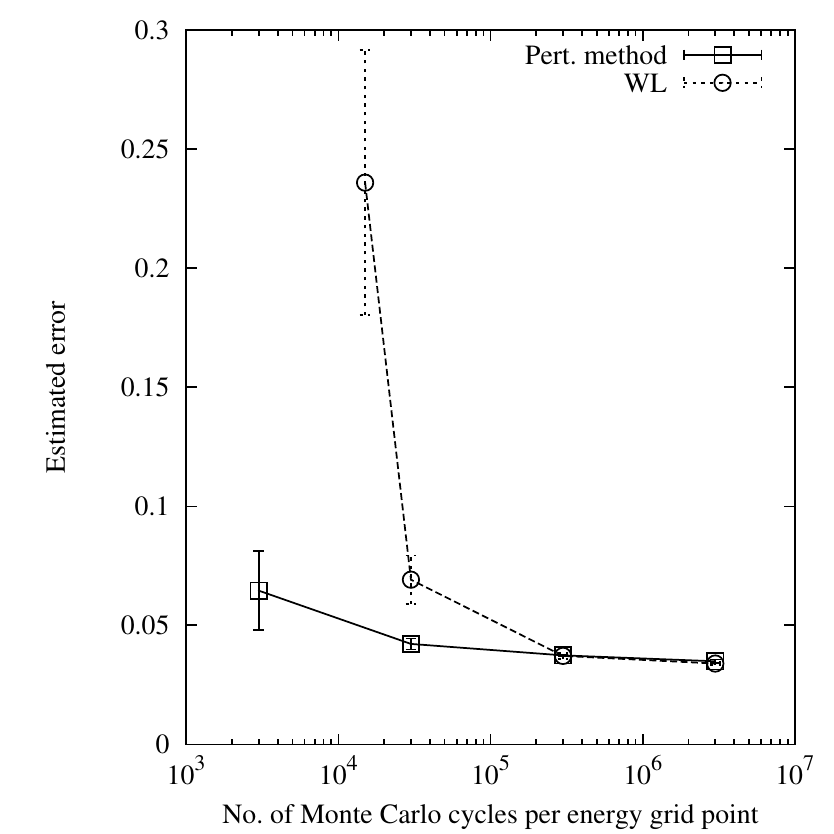}
\caption{Average error ($\langle |\ln \omega_1^\mathrm{est.}(E) - \ln
\omega_1^\mathrm{exact}(E)|\rangle$) as a function of total Monte Carlo cycles
in the energy interval $E \in [50.4, 133]$ (with respect to the ground-state
energy). Each point is the average of three independent runs.}
\label{fig:swerr}
\end{figure}

In Figure~\ref{fig:swerr}, we see the level of convergence attained as a
function of the Monte Carlo cycles for both the perturbation calculations and
the WL sampling for the ``low-medium'' density. The implementation of the WL
sampling is in everything essential the same as for the oscillators discussed
earlier. The maximum displacement in the random walk was the same as for the
perturbation calculations, $1.0$ units in the configurational space and the
same in momentum space (the mass being taken as unity). Like then, the error
was estimated by comparing the slope of the partially converged $\ln \omega(E)$
of one method, to that of the converged $\ln \omega(E)$ of the other method.
Because $\ln \omega(E)$ is not a linear function in $E$, and for the reasons
discussed earlier, perfect agreement between the two converged derivatives is
not attained with the numerical differentiation. The apparent limiting error is
about $0.03$ for the average unsigned difference between the two calculated
slopes, which is sufficient for the comparison.  

Like in the case of the oscillators, it is clear that the perturbation
calculations exhibit a greater degree of convergence already after a small
number of cycles than the WL sampling. When the two algorithms are close to
maximum convergence, they become more difficult to distinguish. We also note
that although not apparent in this calculation---because the calculation of the
energy of the square-well fluid is computationally trivial---the number of
energy evaluations for the perturbation calculations only constitute $1/20$:th
of the total number of cycles (this is because 95\% of the cycles are
arbitrarily dedicated to propagating the hard-sphere Markov chain). This will
have an important speed impact when considering more demanding interaction
potentials, \textit{e. g.}, many-body potentials. One of the most efficient
cases for the perturbation method would thus seem to be the calculation of the
DOS of many-body potentials with hard-cores (so that the hard-sphere reference
system can be used with benefit).

\subsection{Liquid gold}
\label{sec:gold}
So far, we have only considered systems of low dimensionality and simple
Hamiltonians. This has allowed us to compare the speed and accuracy with the WL
algorithm, and the calculation of heat capacity with canonical Monte Carlo
simulations, at no excessive numerical demands. However, the method is also
applicable to higher dimensions and more demanding Hamiltonians, provided there
is a suitable higher-dimensional reference system to use. When there is, the
full benefits of the method are realized. However, the method may always be
applied using the ideal gas as the reference system. Although the full power of
the method relative to other approaches is not realized (because the overlap
between the system of interest and the reference is small), it is always
possible in principle to carry out the calculation. To illustrate this, our
final example is the calculation of the vapor pressure of liquid gold. We will
consider a $N = 108$-particle system with periodic boundary conditions.

\subsubsection{Numerical protocol}
For completeness, we note that the normalized DOS of the reference ideal gas
system is given by
\begin{equation}
\omega_0(E) = \frac {(2 \pi m)^{3 N / 2} V^N} {h^{3 N} N! \Gamma(3 N / 2)} E^{3
N / 2 - 1},
\end{equation}
where $V$ is the volume, $m$ the particle mass and $\Gamma(x)$ denotes the
Euler $\Gamma$-function. Here we have included the center-of-mass motion as one
of the degrees of freedom. This is the natural result of our Monte Carlo
approach. In molecular dynamics implementations, that would not be
the case, and consequently the reference DOS would be slightly different
\cite{lado81, ray99, shirts06}. This needs to be kept in mind if a potentially
more efficient molecular dynamics sampling is to be attempted. Since
$\omega_0(E)$ is known completely, we make use of it in conjunction with
eq.~(\ref{eq:selfsimilar}) to obtain $\omega_1(E)$ at any $E$ from simulations
over different $\lambda$ at a single $E$.

The gold metal was described by the many-body Sutton-Chen (SC) type potential
\cite{sutton90},
\begin{equation}
U_\mathrm{tot} = \sum_{i=1}^{N} \left [ \sum_{j > i}^N \epsilon \left (\frac
{a} {r_{ij}} \right )^n - c \epsilon \sqrt{\sum_{j \neq i}^N \left (\frac {a}
{r_{ij}} \right )^m} \right ],
\end{equation}
where the parameters $n, m, a, \epsilon, c$ are taken from the empirical
parametrization of \c{C}agin \emph{et al.} \cite{cagin99} intended for
classical simulations. The values for Au are $n = 11$, $m = 8$, $\epsilon =
7.8863 \times 10^{-3}$ eV, $a = 4.0651$ \AA\ and $c = 53.082$. Because of the
much extra numerical work required for the $N = 108$ system and the many-body
potential, the calculations to be reported have been obtained from eight
independent ``Intel Xeon E5520'' 2.27 GHz processor cores on a parallel computer
architecture.\footnote{On such cores, what takes 200 CPU minutes
on the author's laptop takes around 80. On another laptop
with a 2.50 GHz ``Intel Core i5-2450'' CPU, this reduces to 45 minutes.}
A single $(E, \lambda)$-point took about two hours of processor time when run
for $10^7$ Monte Carlo cycles and this was deemed acceptable accuracy. The
$\lambda$-parameter was scaled from $1.0$ to $0.05$ in steps of $\Delta \lambda
= 0.05$ at $E = 20$ eV, and from thereon in successive halvings until $\lambda
= 0.0000244140625$; the last point at $\lambda = 0.0$ was calculated by
extrapolation (\emph{vide infra}).  All simulations were carried out at a
density of $17.29$ g / cm$^3$, corresponding according to Paradis and coworkers
\cite{paradis08} to the average liquid density in the temperature range
1337--1500 K for their recent density measurements, for which the thermal
volume expansion is about 1\% over the same temperature interval.  Therefore, a
further simplification we make is that the thermal expansion coefficient of our
system is taken to be zero.  Considering the simple (in relation to the ``real
world'') interaction potential, this approximation seems justifiable. The
energy minimum was taken as the single-point energy of the fcc symmetry at this
density, and was $U_0 = -399.08$ eV. It is generally not crucial to have an
exact value of the potential energy minimum, as an error in this quantity will
primarily affect the DOS at the low end of the energy range, which translates
to low temperatures in the partition function, corresponding to the crystalline
state.

\subsubsection{Results}
\begin{figure}
\includegraphics{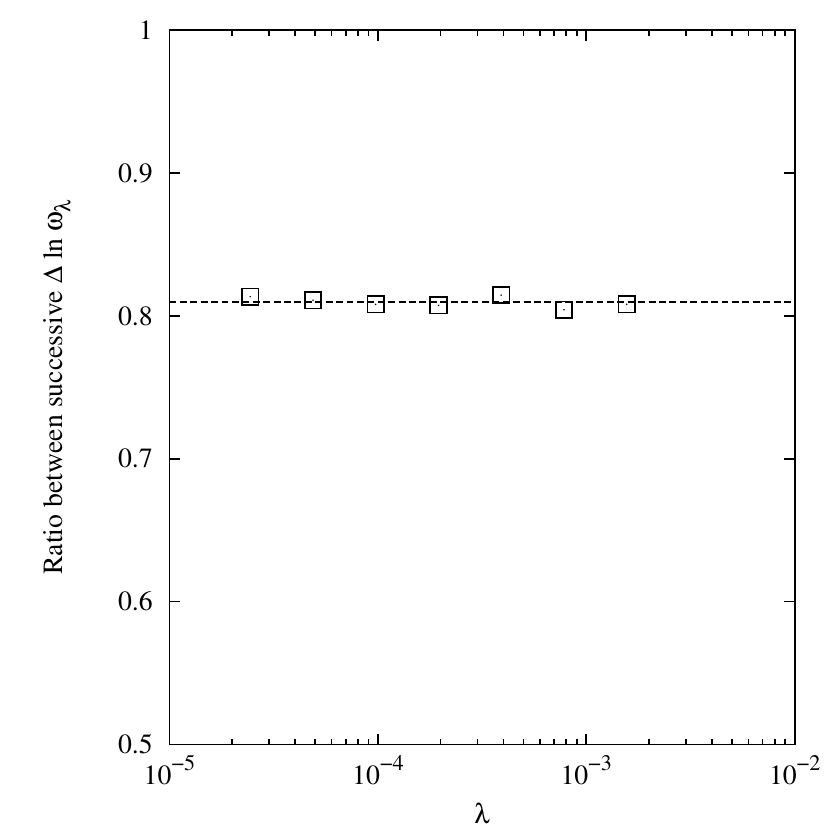}
\caption{Ratio between successive increments $\Delta \ln \omega_\lambda (E)$
for each halving of the $\lambda$-value in the region $\lambda \leq 0.0015625$.
The dashed line is the average value ($0.810 \pm 0.001$) used in the
extrapolation.}
\label{fig:extrapol}
\end{figure}

\begin{figure}
\includegraphics{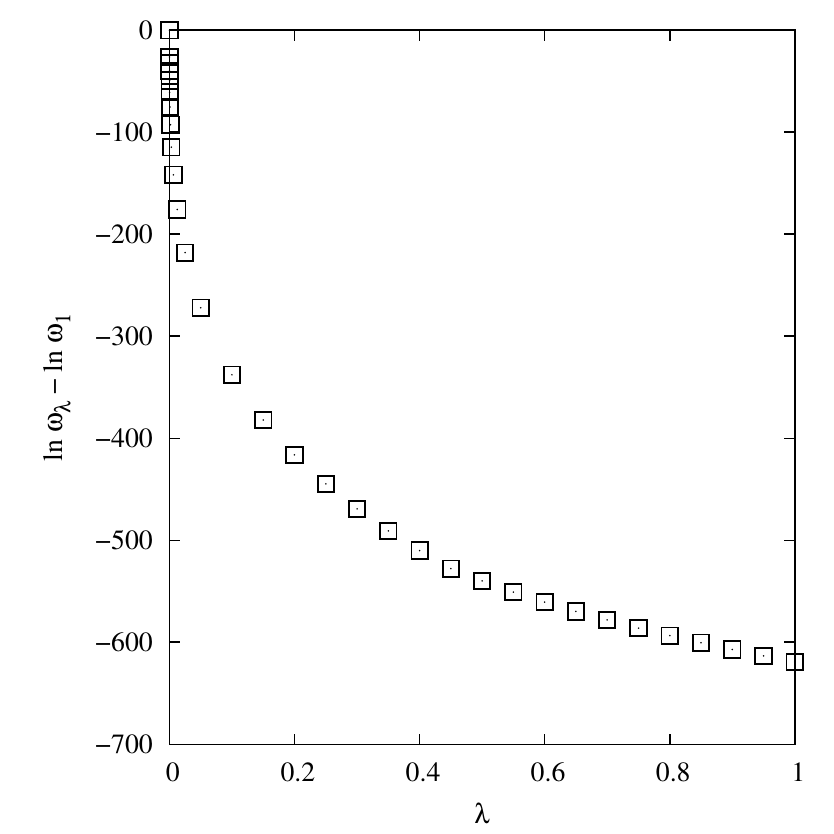}
\caption{Difference between the logarithm of $\omega_{\lambda}(E)$ and
$\omega_{0}(E)$ in the calculation of the DOS of liquid gold for $E = 20$ eV
and $N = 108$. From this curve any arbitrary $E$-point of $\omega_1(E)$ is
obtainable through eq.~(\ref{eq:selfsimilar}).}
\label{fig:logom}
\end{figure}

The short-range repulsion of the interatomic potential is very steep and
resilient to the linear $\lambda$-scaling. Connecting with the point at
$\lambda = 0$ furthermore would seem to require unbiased random sampling, as
the states of the ideal gas are completely random. This step is analogous to
the first energy partitioning window in the NS method, which is also obtained
by random sampling. Random sampling is inefficient. However, when $\lambda$ is
scaled in exponential fashion in the region close to zero, a clear trend
is visible (Figure~\ref{fig:extrapol}) which allows us to extrapolate to
$\lambda = 0$ by the geometric series. The resulting curve of $\ln
\omega_\lambda(E)$ as a function of $\lambda$ is shown in
Figure~\ref{fig:logom}. The extrapolated part represents about 5\% of the
cumulative total value.

The vapor pressure was calculated according to, 
\begin{equation}
p_\mathrm{vap}(T) = \frac {k T} {\Lambda^{3}} \frac
{e^{U_0 / N k T + 1}} {(2 \pi N)^{1 / 2N} Q_\mathrm{Au}^{1/N}},
\end{equation}
where $Q_\mathrm{Au}$ is the partition function of the gold metal. This
equation is derived in the Appendix. We see in Figure
\ref{fig:goldpvap} the calculated vapor pressure as a function of temperature,
compared with experimental estimates \cite{stull47}. When judging the quality of
the results, it must be kept in mind that the SC potential model is a very
simple representation, and the parametrization employed has been derived from
properties of the crystalline, and not the liquid, metal. The potential is
clearly not perfect as, for instance, the relative error in the predicted
surface tension well exceeds 50\% \cite{cagin99}. It should come as no surprise
then, that the absolute value of the predicted vapor pressure is off by roughly
a factor of $3.8$--$4.2$ over the temperature interval considered, with the
slightly better agreement at the high end of the range. The variation in this
factor of around 10\% is smaller than the absolute error, and if the results
are interpreted physically in terms of the Clausius-Clapeyron equation,
\begin{equation}
\ln \frac {p_\mathrm{vap}} {p_0} = -\frac {\Delta_\mathrm{vap} \overline H} {k
T} + \frac {\Delta_\mathrm{vap} \overline S} {k},
\end{equation}
where $p_0$ is the pressure of some reference state (its definition is
arbitrary but affects the value of $\Delta_\mathrm{vap} \overline S$), we see
that this accuracy of the slope translates into a good estimate of the
molecular enthalpy of vaporization, $\Delta_\mathrm{vap} \overline H$. It is
hence primarily the molecular entropy of vaporization, $\Delta_\mathrm{vap}
\overline S$, which is underestimated by this parametrization of the SC
potential. It is not surprising that the accuracy in $\Delta_\mathrm{vap}
\overline H$ is higher, as it is related to the average well-depth of the
interatomic potential, and has been explicitly fitted for the crystal.
$\Delta_\mathrm{vap} \overline S$, on the other hand, is related to the
\emph{shape} of the interatomic potential and is a much more difficult quantity
to parametrize. 

\begin{figure}
\includegraphics{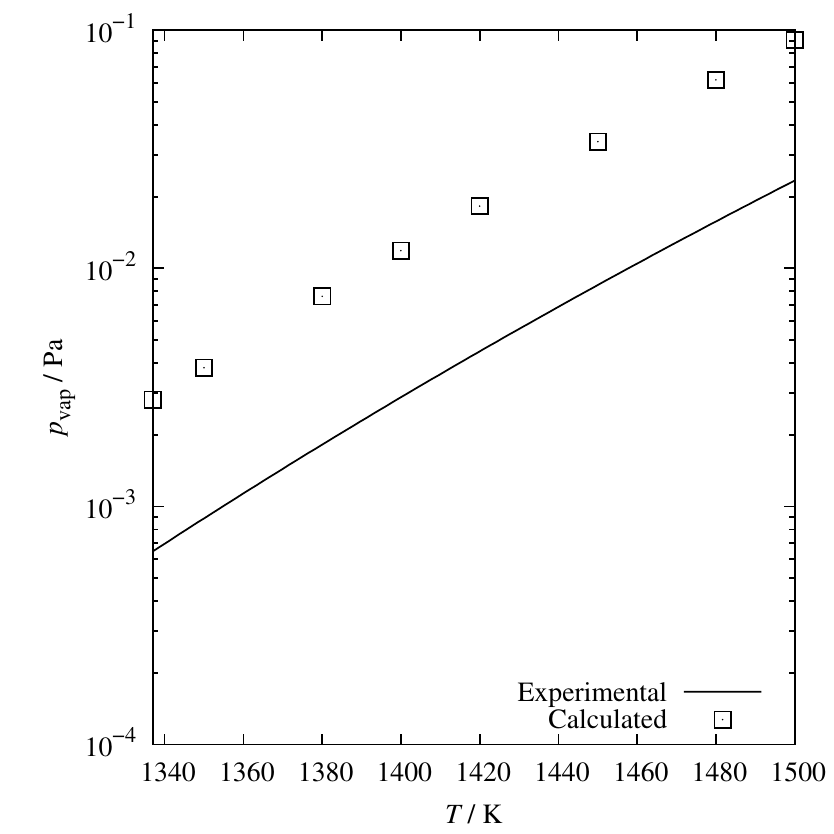}
\caption{Temperature dependence of the calculated and experimental vapor
pressure of gold. The experimental curve is from a fit to the Antoine equation
in the temperature interval $2141$--$3239$ K and is an extrapolation over the
temperature interval considered here.}
\label{fig:goldpvap}
\end{figure}

\section{Conclusion}
In this Paper, it has been shown that calculating the CPF through the DOS by a
perturbation method is a viable alternative to other techniques if the DOS of a
related system is known. The present alternative was found faster than the WL
method for the three-dimensional anharmonic Einstein crystal, and for the
square-well tetradecamer. Technically, the algorithm amounts to sampling (at
most) two microcanonical ensemble averages and so must be considered very
simple. Indeed, one would only need to add a couple of lines of code to
pre-existing molecular dynamics programs, for instance, to implement this
algorithm; and it would require also but very modest modifications to most
Monte Carlo programs to implement the microcanonical average. The greatest
obstacle to a pain-free
implementation of this method is that the potential energy minimum value has to
be independent of $\lambda$, requiring at the very least that efficient energy
minimization can be carried out on the systems of interest. However, a poor
determination of the energy minimum will affect the low-energy region of the
DOS disproportionally, and so a very precise determination might not be
necessary if one is interested in the high-energy end. Another mitigating
factor is the obvious fact that for any method or algorithm to calculate the
low-energy DOS, such energy minimization must be carried out implicitly.
Systems for which energy minimization is difficult, for whatever reason, are
thus inherently difficult cases for which to calculate the complete DOS by any
method. Incidentally, we note that efficient energy minimization is also a
prerequisite of the WL-like algorithm of Soudan \emph{et al.} \cite{soudan11}.

The foremost advantage of the method is that to calculate the DOS of a system
similar to one for which this quantity is already
known, the least possible extra numerical expenditure should be necessary.
However, the greatest drawback of the method is that prior knowledge of the
DOS is generally very scarce. This limits the optimal applicability of this
method because the repertoire of systems with known DOS does not necessarily
include those that are related to the system of study. It is therefore
foreseeable that this algorithm will be most useful in conjunction with another
method to calculate the DOS. Like this, once obtained for one system, a whole
series of related systems will be amenable to structured investigation. Such a
combination of methods could be, for instance, ``WL plus perturbation'' or
a similar recipe. The cost of acquiring the DOS of the reference system, by
whatever suitable method, is then offset by the ease of calculation of the DOS
of the related systems.  Also, unless the absolute DOS is needed (to compute,
for instance, a phase diagram) in some situations entropic \emph{differences}
may suffice. 

However, one additional advantage of the perturbation method is its ability to
calculate $\omega(E)$ at any $E$-value, independently of the $E$-range one
ultimately considers, which means that the DOS can be gradually accrued from
completely separate simulations without the need of having to decide on a
discretization scheme beforehand. This means that the algorithm is trivially
parallelizable and also opens up a vast array of possibilities for further
improvement. For instance, a ``smart,'' {\em e. g.} automatic and non-uniform,
discretization of the energy levels when calculating the DOS, so that those
regions where the DOS varies most rapidly are sampled most thoroughly, is a
natural extension, somewhat analogous to the energy segment partitioning of the
NS method. 

\begin{acknowledgments}
Use of the computer resources of the Chalmers Centre for Computational Science
and Engineering (C$^3$SE) under project SNIC001-11-280 is gratefully
acknowledged. I am thankful to Prof. Roland Kjellander, and to anonymous
Referees, for many constructive comments and criticisms.
\end{acknowledgments}

\appendix*
\section{Derivation of the gold vapor pressure equation}
In the one-component system that we consider, the chemical potential of the
liquid is related to the Helmholtz free energy $A_\mathrm{Au} = U_0 -k T \ln
Q_\mathrm{Au}$ through $\mu_\mathrm{Au} = (A_\mathrm{Au} + p_\mathrm{vap}
V) / N$. Experimentally, the product $p_\mathrm{vap} V / N$ is around
$10^{-30}$ J whereas the calculated $A_\mathrm{Au} / N$ is around $10^{-18}$ J.
The second term may therefore safely be neglected in view of the other
approximations involved. The chemical potential of the vapor is
$\mu_\mathrm{vap} = -k T \partial \ln Q_\mathrm{vap} / \partial N$. Setting
$\mu_\mathrm{Au} = \mu_\mathrm{vap}$, neglecting the pressure-volume term and
substituting $Q_\mathrm{vap} = V^N / (\Lambda^{3 N} N!)$, one arrives at
the result quoted in the text after applying Stirling's approximation, $N!
\approx \sqrt{2 \pi N} N^N e^{-N}$.

%\bibliography{patron}

\begin{thebibliography}{59}%
\makeatletter
\providecommand \@ifxundefined [1]{%
 \@ifx{#1\undefined}
}%
\providecommand \@ifnum [1]{%
 \ifnum #1\expandafter \@firstoftwo
 \else \expandafter \@secondoftwo
 \fi
}%
\providecommand \@ifx [1]{%
 \ifx #1\expandafter \@firstoftwo
 \else \expandafter \@secondoftwo
 \fi
}%
\providecommand \natexlab [1]{#1}%
\providecommand \enquote  [1]{``#1''}%
\providecommand \bibnamefont  [1]{#1}%
\providecommand \bibfnamefont [1]{#1}%
\providecommand \citenamefont [1]{#1}%
\providecommand \href@noop [0]{\@secondoftwo}%
\providecommand \href [0]{\begingroup \@sanitize@url \@href}%
\providecommand \@href[1]{\@@startlink{#1}\@@href}%
\providecommand \@@href[1]{\endgroup#1\@@endlink}%
\providecommand \@sanitize@url [0]{\catcode `\\12\catcode `\$12\catcode
  `\&12\catcode `\#12\catcode `\^12\catcode `\_12\catcode `\%12\relax}%
\providecommand \@@startlink[1]{}%
\providecommand \@@endlink[0]{}%
\providecommand \url  [0]{\begingroup\@sanitize@url \@url }%
\providecommand \@url [1]{\endgroup\@href {#1}{\urlprefix }}%
\providecommand \urlprefix  [0]{URL }%
\providecommand \Eprint [0]{\href }%
\providecommand \doibase [0]{http://dx.doi.org/}%
\providecommand \selectlanguage [0]{\@gobble}%
\providecommand \bibinfo  [0]{\@secondoftwo}%
\providecommand \bibfield  [0]{\@secondoftwo}%
\providecommand \translation [1]{[#1]}%
\providecommand \BibitemOpen [0]{}%
\providecommand \bibitemStop [0]{}%
\providecommand \bibitemNoStop [0]{.\EOS\space}%
\providecommand \EOS [0]{\spacefactor3000\relax}%
\providecommand \BibitemShut  [1]{\csname bibitem#1\endcsname}%
\let\auto@bib@innerbib\@empty
%</preamble>
\bibitem [{\citenamefont {Dunkel}\ and\ \citenamefont
  {Hilbert}(2006)}]{dunkel06}%
  \BibitemOpen
  \bibfield  {author} {\bibinfo {author} {\bibfnamefont {J.}~\bibnamefont
  {Dunkel}}\ and\ \bibinfo {author} {\bibfnamefont {S.}~\bibnamefont
  {Hilbert}},\ }\href {\doibase {10.1016/j.physa.2006.05.018}} {\bibfield
  {journal} {\bibinfo  {journal} {{Physica A}}\ }\textbf {\bibinfo {volume}
  {{370}}},\ \bibinfo {pages} {390} (\bibinfo {year} {{2006}})}\BibitemShut
  {NoStop}%
\bibitem [{\citenamefont {Ming}\ \emph {et~al.}(1996)\citenamefont {Ming},
  \citenamefont {Nordholm},\ and\ \citenamefont {Schranz}}]{ming96}%
  \BibitemOpen
  \bibfield  {author} {\bibinfo {author} {\bibfnamefont {L.}~\bibnamefont
  {Ming}}, \bibinfo {author} {\bibfnamefont {S.}~\bibnamefont {Nordholm}}, \
  and\ \bibinfo {author} {\bibfnamefont {H.~W.}\ \bibnamefont {Schranz}},\
  }\href@noop {} {\bibfield  {journal} {\bibinfo  {journal} {{Chem. Phys.
  Lett.}}\ }\textbf {\bibinfo {volume} {{248}}},\ \bibinfo {pages} {228}
  (\bibinfo {year} {{1996}})}\BibitemShut {NoStop}%
\bibitem [{\citenamefont {Labastie}\ and\ \citenamefont
  {Whetten}(1990)}]{labastie90}%
  \BibitemOpen
  \bibfield  {author} {\bibinfo {author} {\bibfnamefont {P.}~\bibnamefont
  {Labastie}}\ and\ \bibinfo {author} {\bibfnamefont {R.~L.}\ \bibnamefont
  {Whetten}},\ }\href {\doibase {10.1103/PhysRevLett.65.1567}} {\bibfield
  {journal} {\bibinfo  {journal} {{Phys. Rev. Lett.}}\ }\textbf {\bibinfo
  {volume} {{65}}},\ \bibinfo {pages} {1567} (\bibinfo {year}
  {{1990}})}\BibitemShut {NoStop}%
\bibitem [{\citenamefont {Cheng}\ \emph {et~al.}(1992)\citenamefont {Cheng},
  \citenamefont {Li}, \citenamefont {Whetten},\ and\ \citenamefont
  {Berry}}]{cheng92}%
  \BibitemOpen
  \bibfield  {author} {\bibinfo {author} {\bibfnamefont {H.-P.}\ \bibnamefont
  {Cheng}}, \bibinfo {author} {\bibfnamefont {X.}~\bibnamefont {Li}}, \bibinfo
  {author} {\bibfnamefont {R.~L.}\ \bibnamefont {Whetten}}, \ and\ \bibinfo
  {author} {\bibfnamefont {R.~S.}\ \bibnamefont {Berry}},\ }\href {\doibase
  {10.1103/PhysRevA.46.791}} {\bibfield  {journal} {\bibinfo  {journal} {{Phys.
  Rev. A}}\ }\textbf {\bibinfo {volume} {{46}}},\ \bibinfo {pages} {791}
  (\bibinfo {year} {{1992}})}\BibitemShut {NoStop}%
\bibitem [{\citenamefont {Poteau}\ \emph {et~al.}(1994)\citenamefont {Poteau},
  \citenamefont {Spiegelmann},\ and\ \citenamefont {Labastie}}]{poteau94}%
  \BibitemOpen
  \bibfield  {author} {\bibinfo {author} {\bibfnamefont {R.}~\bibnamefont
  {Poteau}}, \bibinfo {author} {\bibfnamefont {F.}~\bibnamefont {Spiegelmann}},
  \ and\ \bibinfo {author} {\bibfnamefont {P.}~\bibnamefont {Labastie}},\
  }\href {\doibase {10.1007/BF01437480}} {\bibfield  {journal} {\bibinfo
  {journal} {{Z. Phys. D Atom. Mol. Clu.}}\ }\textbf {\bibinfo {volume}
  {{30}}},\ \bibinfo {pages} {57} (\bibinfo {year} {{1994}})}\BibitemShut
  {NoStop}%
\bibitem [{\citenamefont {Calvo}\ and\ \citenamefont
  {Labastie}(1995)}]{calvo95}%
  \BibitemOpen
  \bibfield  {author} {\bibinfo {author} {\bibfnamefont {F.}~\bibnamefont
  {Calvo}}\ and\ \bibinfo {author} {\bibfnamefont {P.}~\bibnamefont
  {Labastie}},\ }\href {\doibase {10.1016/S0009-2614(95)01226-5}} {\bibfield
  {journal} {\bibinfo  {journal} {{Chem. Phys. Lett.}}\ }\textbf {\bibinfo
  {volume} {{247}}},\ \bibinfo {pages} {395} (\bibinfo {year}
  {{1995}})}\BibitemShut {NoStop}%
\bibitem [{\citenamefont {Ferrenberg}\ and\ \citenamefont
  {Swendsen}(1988)}]{ferrenberg88}%
  \BibitemOpen
  \bibfield  {author} {\bibinfo {author} {\bibfnamefont {A.~M.}\ \bibnamefont
  {Ferrenberg}}\ and\ \bibinfo {author} {\bibfnamefont {R.~H.}\ \bibnamefont
  {Swendsen}},\ }\href@noop {} {\bibfield  {journal} {\bibinfo  {journal}
  {{Phys. Rev. Lett.}}\ }\textbf {\bibinfo {volume} {{61}}},\ \bibinfo {pages}
  {2635} (\bibinfo {year} {{1988}})}\BibitemShut {NoStop}%
\bibitem [{\citenamefont {Ferrenberg}\ and\ \citenamefont
  {Swendsen}(1989)}]{ferrenberg89}%
  \BibitemOpen
  \bibfield  {author} {\bibinfo {author} {\bibfnamefont {A.~M.}\ \bibnamefont
  {Ferrenberg}}\ and\ \bibinfo {author} {\bibfnamefont {R.~H.}\ \bibnamefont
  {Swendsen}},\ }\href {\doibase {10.1103/PhysRevLett.63.1195}} {\bibfield
  {journal} {\bibinfo  {journal} {{Phys. Rev. Lett.}}\ }\textbf {\bibinfo
  {volume} {{63}}},\ \bibinfo {pages} {1195} (\bibinfo {year}
  {{1989}})}\BibitemShut {NoStop}%
\bibitem [{\citenamefont {Wang}\ and\ \citenamefont
  {Landau}(2001{\natexlab{a}})}]{wang01a}%
  \BibitemOpen
  \bibfield  {author} {\bibinfo {author} {\bibfnamefont {F.}~\bibnamefont
  {Wang}}\ and\ \bibinfo {author} {\bibfnamefont {D.~P.}\ \bibnamefont
  {Landau}},\ }\href {\doibase {10.1103/PhysRevLett.86.2050}} {\bibfield
  {journal} {\bibinfo  {journal} {{Phys. Rev. Lett.}}\ }\textbf {\bibinfo
  {volume} {{86}}},\ \bibinfo {pages} {2050} (\bibinfo {year}
  {{2001}}{\natexlab{a}})}\BibitemShut {NoStop}%
\bibitem [{\citenamefont {Wang}\ and\ \citenamefont
  {Landau}(2001{\natexlab{b}})}]{wang01b}%
  \BibitemOpen
  \bibfield  {author} {\bibinfo {author} {\bibfnamefont {F.}~\bibnamefont
  {Wang}}\ and\ \bibinfo {author} {\bibfnamefont {D.~P.}\ \bibnamefont
  {Landau}},\ }\href {\doibase {10.1103/PhysRevE.64.056101}} {\bibfield
  {journal} {\bibinfo  {journal} {{Phys. Rev. E}}\ }\textbf {\bibinfo {volume}
  {{64}}},\ \bibinfo {pages} {056101} (\bibinfo {year}
  {{2001}}{\natexlab{b}})}\BibitemShut {NoStop}%
\bibitem [{\citenamefont {Berg}\ and\ \citenamefont {Neuhaus}(1991)}]{berg91}%
  \BibitemOpen
  \bibfield  {author} {\bibinfo {author} {\bibfnamefont {B.~A.}\ \bibnamefont
  {Berg}}\ and\ \bibinfo {author} {\bibfnamefont {T.}~\bibnamefont {Neuhaus}},\
  }\href {\doibase {10.1016/0370-2693(91)91256-U}} {\bibfield  {journal}
  {\bibinfo  {journal} {{Phys. Lett. B}}\ }\textbf {\bibinfo {volume}
  {{267}}},\ \bibinfo {pages} {249} (\bibinfo {year} {{1991}})}\BibitemShut
  {NoStop}%
\bibitem [{\citenamefont {Lee}(1993)}]{lee93}%
  \BibitemOpen
  \bibfield  {author} {\bibinfo {author} {\bibfnamefont {J.}~\bibnamefont
  {Lee}},\ }\href {\doibase {10.1103/PhysRevLett.71.211}} {\bibfield  {journal}
  {\bibinfo  {journal} {{Phys. Rev. Lett.}}\ }\textbf {\bibinfo {volume}
  {{71}}},\ \bibinfo {pages} {211} (\bibinfo {year} {{1993}})}\BibitemShut
  {NoStop}%
\bibitem [{\citenamefont {Geyer}\ and\ \citenamefont
  {Thompson}(1995)}]{geyer95}%
  \BibitemOpen
  \bibfield  {author} {\bibinfo {author} {\bibfnamefont {C.~J.}\ \bibnamefont
  {Geyer}}\ and\ \bibinfo {author} {\bibfnamefont {E.~A.}\ \bibnamefont
  {Thompson}},\ }\href@noop {} {\bibfield  {journal} {\bibinfo  {journal} {{J.
  Am. Stat. Assoc.}}\ }\textbf {\bibinfo {volume} {{90}}},\ \bibinfo {pages}
  {909} (\bibinfo {year} {{1995}})}\BibitemShut {NoStop}%
\bibitem [{\citenamefont {Wang}\ \emph {et~al.}(1999)\citenamefont {Wang},
  \citenamefont {Tay},\ and\ \citenamefont {Swendsen}}]{wang99}%
  \BibitemOpen
  \bibfield  {author} {\bibinfo {author} {\bibfnamefont {J.-S.}\ \bibnamefont
  {Wang}}, \bibinfo {author} {\bibfnamefont {T.~K.}\ \bibnamefont {Tay}}, \
  and\ \bibinfo {author} {\bibfnamefont {R.~H.}\ \bibnamefont {Swendsen}},\
  }\href {\doibase {10.1103/PhysRevLett.82.476}} {\bibfield  {journal}
  {\bibinfo  {journal} {{Phys. Rev. Lett.}}\ }\textbf {\bibinfo {volume}
  {{82}}},\ \bibinfo {pages} {476} (\bibinfo {year} {{1999}})}\BibitemShut
  {NoStop}%
\bibitem [{\citenamefont {Heilmann}\ and\ \citenamefont
  {Hoffmann}(2005)}]{heilmann05}%
  \BibitemOpen
  \bibfield  {author} {\bibinfo {author} {\bibfnamefont {F.}~\bibnamefont
  {Heilmann}}\ and\ \bibinfo {author} {\bibfnamefont {K.~H.}\ \bibnamefont
  {Hoffmann}},\ }\href {\doibase 10.1209/epl/i2004-10486-9} {\bibfield
  {journal} {\bibinfo  {journal} {Europhys. Lett.}\ }\textbf {\bibinfo {volume}
  {70}},\ \bibinfo {pages} {155} (\bibinfo {year} {2005})}\BibitemShut
  {NoStop}%
\bibitem [{\citenamefont {Skilling}(2004)}]{skilling04}%
  \BibitemOpen
  \bibfield  {author} {\bibinfo {author} {\bibfnamefont {J.}~\bibnamefont
  {Skilling}},\ }in\ \href@noop {} {\emph {\bibinfo {booktitle} {{AIP Conf.
  Proc.}}}},\ Vol.\ \bibinfo {volume} {{735}}\ (\bibinfo {year} {{2004}})\ p.\
  \bibinfo {pages} {395}\BibitemShut {NoStop}%
\bibitem [{\citenamefont {Skilling}(2006)}]{skilling06}%
  \BibitemOpen
  \bibfield  {author} {\bibinfo {author} {\bibfnamefont {J.}~\bibnamefont
  {Skilling}},\ }\href@noop {} {\bibfield  {journal} {\bibinfo  {journal}
  {{Bayesian Anal.}}\ }\textbf {\bibinfo {volume} {{1}}},\ \bibinfo {pages}
  {833} (\bibinfo {year} {{2006}})}\BibitemShut {NoStop}%
\bibitem [{\citenamefont {P{\'{a}}rtay}\ \emph {et~al.}(2010)\citenamefont
  {P{\'{a}}rtay}, \citenamefont {Bart{\'{o}}k},\ and\ \citenamefont
  {Cs{\'{a}}nyi}}]{partay10}%
  \BibitemOpen
  \bibfield  {author} {\bibinfo {author} {\bibfnamefont {L.~B.}\ \bibnamefont
  {P{\'{a}}rtay}}, \bibinfo {author} {\bibfnamefont {A.~P.}\ \bibnamefont
  {Bart{\'{o}}k}}, \ and\ \bibinfo {author} {\bibfnamefont {G.}~\bibnamefont
  {Cs{\'{a}}nyi}},\ }\href@noop {} {\bibfield  {journal} {\bibinfo  {journal}
  {{J. Phys. Chem. B}}\ }\textbf {\bibinfo {volume} {{114}}},\ \bibinfo {pages}
  {10502} (\bibinfo {year} {{2010}})}\BibitemShut {NoStop}%
\bibitem [{\citenamefont {Do}\ \emph {et~al.}(2011)\citenamefont {Do},
  \citenamefont {Hirst},\ and\ \citenamefont {Wheatley}}]{do11}%
  \BibitemOpen
  \bibfield  {author} {\bibinfo {author} {\bibfnamefont {H.}~\bibnamefont
  {Do}}, \bibinfo {author} {\bibfnamefont {J.~D.}\ \bibnamefont {Hirst}}, \
  and\ \bibinfo {author} {\bibfnamefont {R.~J.}\ \bibnamefont {Wheatley}},\
  }\href@noop {} {\bibfield  {journal} {\bibinfo  {journal} {{J. Chem. Phys.}}\
  }\textbf {\bibinfo {volume} {{135}}},\ \bibinfo {pages} {174105} (\bibinfo
  {year} {{2011}})}\BibitemShut {NoStop}%
\bibitem [{\citenamefont {Zwanzig}(1954)}]{zwanzig54}%
  \BibitemOpen
  \bibfield  {author} {\bibinfo {author} {\bibfnamefont {R.~W.}\ \bibnamefont
  {Zwanzig}},\ }\href@noop {} {\bibfield  {journal} {\bibinfo  {journal} {{J.
  Chem. Phys.}}\ }\textbf {\bibinfo {volume} {{22}}},\ \bibinfo {pages} {1420}
  (\bibinfo {year} {{1954}})}\BibitemShut {NoStop}%
\bibitem [{\citenamefont {Hansen}\ and\ \citenamefont
  {Verlet}(1969)}]{hansen69}%
  \BibitemOpen
  \bibfield  {author} {\bibinfo {author} {\bibfnamefont {J.-P.}\ \bibnamefont
  {Hansen}}\ and\ \bibinfo {author} {\bibfnamefont {L.}~\bibnamefont
  {Verlet}},\ }\href {\doibase {10.1103/PhysRev.184.151}} {\bibfield  {journal}
  {\bibinfo  {journal} {{Phys. Rev.}}\ }\textbf {\bibinfo {volume} {{184}}},\
  \bibinfo {pages} {151} (\bibinfo {year} {{1969}})}\BibitemShut {NoStop}%
\bibitem [{\citenamefont {Henderson}\ and\ \citenamefont
  {Barker}(1970)}]{henderson70}%
  \BibitemOpen
  \bibfield  {author} {\bibinfo {author} {\bibfnamefont {D.}~\bibnamefont
  {Henderson}}\ and\ \bibinfo {author} {\bibfnamefont {J.~A.}\ \bibnamefont
  {Barker}},\ }\href {\doibase {10.1103/PhysRevA.1.1266}} {\bibfield  {journal}
  {\bibinfo  {journal} {{Phys. Rev. A}}\ }\textbf {\bibinfo {volume} {{1}}},\
  \bibinfo {pages} {1266} (\bibinfo {year} {{1970}})}\BibitemShut {NoStop}%
\bibitem [{\citenamefont {Torrie}\ and\ \citenamefont
  {Valleau}(1974)}]{torrie74}%
  \BibitemOpen
  \bibfield  {author} {\bibinfo {author} {\bibfnamefont {G.~M.}\ \bibnamefont
  {Torrie}}\ and\ \bibinfo {author} {\bibfnamefont {J.~P.}\ \bibnamefont
  {Valleau}},\ }\href {\doibase {10.1016/0009-2614(74)80109-0}} {\bibfield
  {journal} {\bibinfo  {journal} {{Chem. Phys. Lett.}}\ }\textbf {\bibinfo
  {volume} {{28}}},\ \bibinfo {pages} {578} (\bibinfo {year}
  {{1974}})}\BibitemShut {NoStop}%
\bibitem [{\citenamefont {Torrie}\ and\ \citenamefont
  {Valleau}(1977)}]{torrie77}%
  \BibitemOpen
  \bibfield  {author} {\bibinfo {author} {\bibfnamefont {G.~M.}\ \bibnamefont
  {Torrie}}\ and\ \bibinfo {author} {\bibfnamefont {J.~P.}\ \bibnamefont
  {Valleau}},\ }\href@noop {} {\bibfield  {journal} {\bibinfo  {journal} {{J.
  Comp. Phys.}}\ }\textbf {\bibinfo {volume} {{23}}},\ \bibinfo {pages} {187}
  (\bibinfo {year} {{1977}})}\BibitemShut {NoStop}%
\bibitem [{\citenamefont {Yan}\ \emph {et~al.}(2002)\citenamefont {Yan},
  \citenamefont {Faller},\ and\ \citenamefont {De~Pablo}}]{yan02}%
  \BibitemOpen
  \bibfield  {author} {\bibinfo {author} {\bibfnamefont {Q.}~\bibnamefont
  {Yan}}, \bibinfo {author} {\bibfnamefont {R.}~\bibnamefont {Faller}}, \ and\
  \bibinfo {author} {\bibfnamefont {J.~J.}\ \bibnamefont {De~Pablo}},\
  }\href@noop {} {\bibfield  {journal} {\bibinfo  {journal} {{J. Chem. Phys.}}\
  }\textbf {\bibinfo {volume} {{116}}},\ \bibinfo {pages} {8745} (\bibinfo
  {year} {{2002}})}\BibitemShut {NoStop}%
\bibitem [{\citenamefont {Shell}\ \emph {et~al.}(2002)\citenamefont {Shell},
  \citenamefont {Debenedetti},\ and\ \citenamefont
  {Panagiotopoulos}}]{shell02}%
  \BibitemOpen
  \bibfield  {author} {\bibinfo {author} {\bibfnamefont {M.~S.}\ \bibnamefont
  {Shell}}, \bibinfo {author} {\bibfnamefont {P.~G.}\ \bibnamefont
  {Debenedetti}}, \ and\ \bibinfo {author} {\bibfnamefont {A.~Z.}\ \bibnamefont
  {Panagiotopoulos}},\ }\href@noop {} {\bibfield  {journal} {\bibinfo
  {journal} {{Phys. Rev. E}}\ }\textbf {\bibinfo {volume} {{66}}},\ \bibinfo
  {pages} {056703} (\bibinfo {year} {{2002}})}\BibitemShut {NoStop}%
\bibitem [{\citenamefont {Mauro}\ \emph {et~al.}(2007)\citenamefont {Mauro},
  \citenamefont {Loucks}, \citenamefont {Balakrishnan},\ and\ \citenamefont
  {Raghavan}}]{mauro07}%
  \BibitemOpen
  \bibfield  {author} {\bibinfo {author} {\bibfnamefont {J.}~\bibnamefont
  {Mauro}}, \bibinfo {author} {\bibfnamefont {R.}~\bibnamefont {Loucks}},
  \bibinfo {author} {\bibfnamefont {J.}~\bibnamefont {Balakrishnan}}, \ and\
  \bibinfo {author} {\bibfnamefont {S.}~\bibnamefont {Raghavan}},\ }\href@noop
  {} {\bibfield  {journal} {\bibinfo  {journal} {J. Chem. Phys.}\ }\textbf
  {\bibinfo {volume} {126}},\ \bibinfo {pages} {194103} (\bibinfo {year}
  {2007})}\BibitemShut {NoStop}%
\bibitem [{\citenamefont {Desgranges}\ and\ \citenamefont
  {Delhommelle}(2009)}]{desgranges09}%
  \BibitemOpen
  \bibfield  {author} {\bibinfo {author} {\bibfnamefont {C.}~\bibnamefont
  {Desgranges}}\ and\ \bibinfo {author} {\bibfnamefont {J.}~\bibnamefont
  {Delhommelle}},\ }\href {\doibase 10.1063/1.3158605} {\bibfield  {journal}
  {\bibinfo  {journal} {J. Chem. Phys.}\ }\textbf {\bibinfo {volume} {130}},\
  \bibinfo {eid} {244109} (\bibinfo {year} {2009})}\BibitemShut {NoStop}%
\bibitem [{\citenamefont {Schulz}\ \emph {et~al.}(2002)\citenamefont {Schulz},
  \citenamefont {Binder},\ and\ \citenamefont {M\"uller}}]{schulz02}%
  \BibitemOpen
  \bibfield  {author} {\bibinfo {author} {\bibfnamefont {B.~J.}\ \bibnamefont
  {Schulz}}, \bibinfo {author} {\bibfnamefont {K.}~\bibnamefont {Binder}}, \
  and\ \bibinfo {author} {\bibfnamefont {M.}~\bibnamefont {M\"uller}},\ }\href
  {\doibase 10.1142/S0129183102003243} {\bibfield  {journal} {\bibinfo
  {journal} {Int. J. Mod. Phys. C}\ }\textbf {\bibinfo {volume} {13}},\
  \bibinfo {pages} {477} (\bibinfo {year} {2002})}\BibitemShut {NoStop}%
\bibitem [{\citenamefont {Dayal}\ \emph {et~al.}(2004)\citenamefont {Dayal},
  \citenamefont {Trebst}, \citenamefont {Wessel}, \citenamefont {W\"urtz},
  \citenamefont {Troyer}, \citenamefont {Sabhapandit},\ and\ \citenamefont
  {Coppersmith}}]{dayal04}%
  \BibitemOpen
  \bibfield  {author} {\bibinfo {author} {\bibfnamefont {P.}~\bibnamefont
  {Dayal}}, \bibinfo {author} {\bibfnamefont {S.}~\bibnamefont {Trebst}},
  \bibinfo {author} {\bibfnamefont {S.}~\bibnamefont {Wessel}}, \bibinfo
  {author} {\bibfnamefont {D.}~\bibnamefont {W\"urtz}}, \bibinfo {author}
  {\bibfnamefont {M.}~\bibnamefont {Troyer}}, \bibinfo {author} {\bibfnamefont
  {S.}~\bibnamefont {Sabhapandit}}, \ and\ \bibinfo {author} {\bibfnamefont
  {S.~N.}\ \bibnamefont {Coppersmith}},\ }\href {\doibase
  10.1103/PhysRevLett.92.097201} {\bibfield  {journal} {\bibinfo  {journal}
  {Phys. Rev. Lett.}\ }\textbf {\bibinfo {volume} {92}},\ \bibinfo {pages}
  {097201} (\bibinfo {year} {2004})}\BibitemShut {NoStop}%
\bibitem [{\citenamefont {Tr\"oster}\ and\ \citenamefont
  {Dellago}(2005)}]{troster05}%
  \BibitemOpen
  \bibfield  {author} {\bibinfo {author} {\bibfnamefont {A.}~\bibnamefont
  {Tr\"oster}}\ and\ \bibinfo {author} {\bibfnamefont {C.}~\bibnamefont
  {Dellago}},\ }\href {\doibase 10.1103/PhysRevE.71.066705} {\bibfield
  {journal} {\bibinfo  {journal} {Phys. Rev. E}\ }\textbf {\bibinfo {volume}
  {71}},\ \bibinfo {pages} {066705} (\bibinfo {year} {2005})}\BibitemShut
  {NoStop}%
\bibitem [{\citenamefont {Lee}\ \emph {et~al.}(2006)\citenamefont {Lee},
  \citenamefont {Okabe},\ and\ \citenamefont {Landau}}]{lee06}%
  \BibitemOpen
  \bibfield  {author} {\bibinfo {author} {\bibfnamefont {H.~K.}\ \bibnamefont
  {Lee}}, \bibinfo {author} {\bibfnamefont {Y.}~\bibnamefont {Okabe}}, \ and\
  \bibinfo {author} {\bibfnamefont {D.~P.}\ \bibnamefont {Landau}},\
  }\href@noop {} {\bibfield  {journal} {\bibinfo  {journal} {Comput. Phys.
  Comm.}\ }\textbf {\bibinfo {volume} {175}},\ \bibinfo {pages} {36} (\bibinfo
  {year} {2006})}\BibitemShut {NoStop}%
\bibitem [{\citenamefont {Poulain}\ \emph {et~al.}(2006)\citenamefont
  {Poulain}, \citenamefont {Calvo}, \citenamefont {Antoine}, \citenamefont
  {Broyer},\ and\ \citenamefont {Dugourd}}]{poulain06}%
  \BibitemOpen
  \bibfield  {author} {\bibinfo {author} {\bibfnamefont {P.}~\bibnamefont
  {Poulain}}, \bibinfo {author} {\bibfnamefont {F.}~\bibnamefont {Calvo}},
  \bibinfo {author} {\bibfnamefont {R.}~\bibnamefont {Antoine}}, \bibinfo
  {author} {\bibfnamefont {M.}~\bibnamefont {Broyer}}, \ and\ \bibinfo {author}
  {\bibfnamefont {P.}~\bibnamefont {Dugourd}},\ }\href {\doibase
  10.1103/PhysRevE.73.056704} {\bibfield  {journal} {\bibinfo  {journal} {Phys.
  Rev. E}\ }\textbf {\bibinfo {volume} {73}},\ \bibinfo {pages} {056704}
  (\bibinfo {year} {2006})}\BibitemShut {NoStop}%
\bibitem [{\citenamefont {Belardinelli}\ and\ \citenamefont
  {Pereyra}(2007{\natexlab{a}})}]{belardinelli07b}%
  \BibitemOpen
  \bibfield  {author} {\bibinfo {author} {\bibfnamefont {R.~E.}\ \bibnamefont
  {Belardinelli}}\ and\ \bibinfo {author} {\bibfnamefont {V.~D.}\ \bibnamefont
  {Pereyra}},\ }\href {\doibase 10.1103/PhysRevE.75.046701} {\bibfield
  {journal} {\bibinfo  {journal} {Phys. Rev. E}\ }\textbf {\bibinfo {volume}
  {75}},\ \bibinfo {pages} {046701} (\bibinfo {year}
  {2007}{\natexlab{a}})}\BibitemShut {NoStop}%
\bibitem [{\citenamefont {Zhou}\ and\ \citenamefont {Su}(2008)}]{zhou08}%
  \BibitemOpen
  \bibfield  {author} {\bibinfo {author} {\bibfnamefont {C.}~\bibnamefont
  {Zhou}}\ and\ \bibinfo {author} {\bibfnamefont {J.}~\bibnamefont {Su}},\
  }\href {\doibase 10.1103/PhysRevE.78.046705} {\bibfield  {journal} {\bibinfo
  {journal} {Phys. Rev. E}\ }\textbf {\bibinfo {volume} {78}},\ \bibinfo
  {pages} {046705} (\bibinfo {year} {2008})}\BibitemShut {NoStop}%
\bibitem [{\citenamefont {Cunha-Netto}\ \emph {et~al.}(2008)\citenamefont
  {Cunha-Netto}, \citenamefont {Caparica}, \citenamefont {Tsai}, \citenamefont
  {Dickman},\ and\ \citenamefont {Landau}}]{cunha-netto08}%
  \BibitemOpen
  \bibfield  {author} {\bibinfo {author} {\bibfnamefont {A.~G.}\ \bibnamefont
  {Cunha-Netto}}, \bibinfo {author} {\bibfnamefont {A.~A.}\ \bibnamefont
  {Caparica}}, \bibinfo {author} {\bibfnamefont {S.-H.}\ \bibnamefont {Tsai}},
  \bibinfo {author} {\bibfnamefont {R.}~\bibnamefont {Dickman}}, \ and\
  \bibinfo {author} {\bibfnamefont {D.~P.}\ \bibnamefont {Landau}},\ }\href
  {\doibase 10.1103/PhysRevE.78.055701} {\bibfield  {journal} {\bibinfo
  {journal} {Phys. Rev. E}\ }\textbf {\bibinfo {volume} {78}},\ \bibinfo
  {pages} {055701} (\bibinfo {year} {2008})}\BibitemShut {NoStop}%
\bibitem [{\citenamefont {Soudan}\ \emph {et~al.}(2011)\citenamefont {Soudan},
  \citenamefont {Basire}, \citenamefont {Mestdagh},\ and\ \citenamefont
  {Angeli\'{e}}}]{soudan11}%
  \BibitemOpen
  \bibfield  {author} {\bibinfo {author} {\bibfnamefont {J.-M.}\ \bibnamefont
  {Soudan}}, \bibinfo {author} {\bibfnamefont {M.}~\bibnamefont {Basire}},
  \bibinfo {author} {\bibfnamefont {J.-M.}\ \bibnamefont {Mestdagh}}, \ and\
  \bibinfo {author} {\bibfnamefont {C.}~\bibnamefont {Angeli\'{e}}},\ }\href
  {\doibase 10.1063/1.3647333} {\bibfield  {journal} {\bibinfo  {journal} {J.
  Chem. Phys.}\ }\textbf {\bibinfo {volume} {135}},\ \bibinfo {eid} {144109}
  (\bibinfo {year} {2011})}\BibitemShut {NoStop}%
\bibitem [{\citenamefont {Dickman}\ and\ \citenamefont
  {Cunha-Netto}(2011)}]{dickman11}%
  \BibitemOpen
  \bibfield  {author} {\bibinfo {author} {\bibfnamefont {R.}~\bibnamefont
  {Dickman}}\ and\ \bibinfo {author} {\bibfnamefont {A.~G.}\ \bibnamefont
  {Cunha-Netto}},\ }\href {\doibase 10.1103/PhysRevE.84.026701} {\bibfield
  {journal} {\bibinfo  {journal} {Phys. Rev. E}\ }\textbf {\bibinfo {volume}
  {84}},\ \bibinfo {pages} {026701} (\bibinfo {year} {2011})}\BibitemShut
  {NoStop}%
\bibitem [{\citenamefont {Severin}\ \emph {et~al.}(1978)\citenamefont
  {Severin}, \citenamefont {Freasier}, \citenamefont {Hamer}, \citenamefont
  {Jolly},\ and\ \citenamefont {Nordholm}}]{severin78}%
  \BibitemOpen
  \bibfield  {author} {\bibinfo {author} {\bibfnamefont {E.~S.}\ \bibnamefont
  {Severin}}, \bibinfo {author} {\bibfnamefont {B.~C.}\ \bibnamefont
  {Freasier}}, \bibinfo {author} {\bibfnamefont {N.~D.}\ \bibnamefont {Hamer}},
  \bibinfo {author} {\bibfnamefont {D.~L.}\ \bibnamefont {Jolly}}, \ and\
  \bibinfo {author} {\bibfnamefont {S.}~\bibnamefont {Nordholm}},\ }\href@noop
  {} {\bibfield  {journal} {\bibinfo  {journal} {{Chem. Phys. Lett.}}\ }\textbf
  {\bibinfo {volume} {{57}}},\ \bibinfo {pages} {117} (\bibinfo {year}
  {{1978}})}\BibitemShut {NoStop}%
\bibitem [{\citenamefont {Schranz}\ \emph {et~al.}(1991)\citenamefont
  {Schranz}, \citenamefont {Nordholm},\ and\ \citenamefont
  {Nyman}}]{schranz91}%
  \BibitemOpen
  \bibfield  {author} {\bibinfo {author} {\bibfnamefont {H.~W.}\ \bibnamefont
  {Schranz}}, \bibinfo {author} {\bibfnamefont {S.}~\bibnamefont {Nordholm}}, \
  and\ \bibinfo {author} {\bibfnamefont {G.}~\bibnamefont {Nyman}},\
  }\href@noop {} {\bibfield  {journal} {\bibinfo  {journal} {{J. Chem. Phys.}}\
  }\textbf {\bibinfo {volume} {{94}}},\ \bibinfo {pages} {1487} (\bibinfo
  {year} {{1991}})}\BibitemShut {NoStop}%
\bibitem [{\citenamefont {Ray}(1991)}]{ray91}%
  \BibitemOpen
  \bibfield  {author} {\bibinfo {author} {\bibfnamefont {J.~R.}\ \bibnamefont
  {Ray}},\ }\href {\doibase {10.1103/PhysRevA.44.4061}} {\bibfield  {journal}
  {\bibinfo  {journal} {{Phys. Rev. A}}\ }\textbf {\bibinfo {volume} {{44}}},\
  \bibinfo {pages} {4061} (\bibinfo {year} {{1991}})}\BibitemShut {NoStop}%
\bibitem [{\citenamefont {Bennett}(1976)}]{bennett76}%
  \BibitemOpen
  \bibfield  {author} {\bibinfo {author} {\bibfnamefont {C.~H.}\ \bibnamefont
  {Bennett}},\ }\href@noop {} {\bibfield  {journal} {\bibinfo  {journal} {{J.
  Comput. Phys.}}\ }\textbf {\bibinfo {volume} {{22}}},\ \bibinfo {pages} {245}
  (\bibinfo {year} {{1976}})}\BibitemShut {NoStop}%
\bibitem [{\citenamefont {Lyubartsev}\ \emph {et~al.}(1992)\citenamefont
  {Lyubartsev}, \citenamefont {Martsinovski}, \citenamefont {Shevkunov},\ and\
  \citenamefont {Vorontsov-Velyaminov}}]{lyubartsev92}%
  \BibitemOpen
  \bibfield  {author} {\bibinfo {author} {\bibfnamefont {A.~P.}\ \bibnamefont
  {Lyubartsev}}, \bibinfo {author} {\bibfnamefont {A.~A.}\ \bibnamefont
  {Martsinovski}}, \bibinfo {author} {\bibfnamefont {S.~V.}\ \bibnamefont
  {Shevkunov}}, \ and\ \bibinfo {author} {\bibfnamefont {P.~N.}\ \bibnamefont
  {Vorontsov-Velyaminov}},\ }\href {\doibase {10.1063/1.462133}} {\bibfield
  {journal} {\bibinfo  {journal} {{J. Chem. Phys.}}\ }\textbf {\bibinfo
  {volume} {{96}}},\ \bibinfo {pages} {1776} (\bibinfo {year}
  {{1992}})}\BibitemShut {NoStop}%
\bibitem [{\citenamefont {Schulz}\ \emph {et~al.}(2003)\citenamefont {Schulz},
  \citenamefont {Binder}, \citenamefont {M\"uller},\ and\ \citenamefont
  {Landau}}]{schulz03}%
  \BibitemOpen
  \bibfield  {author} {\bibinfo {author} {\bibfnamefont {B.~J.}\ \bibnamefont
  {Schulz}}, \bibinfo {author} {\bibfnamefont {K.}~\bibnamefont {Binder}},
  \bibinfo {author} {\bibfnamefont {M.}~\bibnamefont {M\"uller}}, \ and\
  \bibinfo {author} {\bibfnamefont {D.~P.}\ \bibnamefont {Landau}},\ }\href
  {\doibase 10.1103/PhysRevE.67.067102} {\bibfield  {journal} {\bibinfo
  {journal} {Phys. Rev. E}\ }\textbf {\bibinfo {volume} {67}},\ \bibinfo
  {pages} {067102} (\bibinfo {year} {2003})}\BibitemShut {NoStop}%
\bibitem [{\citenamefont {Yan}\ and\ \citenamefont {de~Pablo}(2003)}]{yan03}%
  \BibitemOpen
  \bibfield  {author} {\bibinfo {author} {\bibfnamefont {Q.}~\bibnamefont
  {Yan}}\ and\ \bibinfo {author} {\bibfnamefont {J.~J.}\ \bibnamefont
  {de~Pablo}},\ }\href {\doibase 10.1103/PhysRevLett.90.035701} {\bibfield
  {journal} {\bibinfo  {journal} {Phys. Rev. Lett.}\ }\textbf {\bibinfo
  {volume} {90}},\ \bibinfo {pages} {035701} (\bibinfo {year}
  {2003})}\BibitemShut {NoStop}%
\bibitem [{\citenamefont {Belardinelli}\ and\ \citenamefont
  {Pereyra}(2007{\natexlab{b}})}]{belardinelli07a}%
  \BibitemOpen
  \bibfield  {author} {\bibinfo {author} {\bibfnamefont {R.~E.}\ \bibnamefont
  {Belardinelli}}\ and\ \bibinfo {author} {\bibfnamefont {V.~D.}\ \bibnamefont
  {Pereyra}},\ }\href {\doibase 10.1063/1.2803061} {\bibfield  {journal}
  {\bibinfo  {journal} {J. Chem. Phys.}\ }\textbf {\bibinfo {volume} {127}},\
  \bibinfo {eid} {184105} (\bibinfo {year} {2007}{\natexlab{b}})}\BibitemShut
  {NoStop}%
\bibitem [{\citenamefont {Komura}\ and\ \citenamefont
  {Okabe}(2012)}]{komura12}%
  \BibitemOpen
  \bibfield  {author} {\bibinfo {author} {\bibfnamefont {Y.}~\bibnamefont
  {Komura}}\ and\ \bibinfo {author} {\bibfnamefont {Y.}~\bibnamefont {Okabe}},\
  }\href {\doibase 10.1103/PhysRevE.85.010102} {\bibfield  {journal} {\bibinfo
  {journal} {Phys. Rev. E}\ }\textbf {\bibinfo {volume} {85}},\ \bibinfo
  {pages} {010102} (\bibinfo {year} {2012})}\BibitemShut {NoStop}%
\bibitem [{\citenamefont {Morozov}\ and\ \citenamefont
  {Lin}(2007)}]{morozov07}%
  \BibitemOpen
  \bibfield  {author} {\bibinfo {author} {\bibfnamefont {A.~N.}\ \bibnamefont
  {Morozov}}\ and\ \bibinfo {author} {\bibfnamefont {S.~H.}\ \bibnamefont
  {Lin}},\ }\href@noop {} {\bibfield  {journal} {\bibinfo  {journal} {Phys.
  Rev. E}\ }\textbf {\bibinfo {volume} {76}},\ \bibinfo {pages} {026701}
  (\bibinfo {year} {2007})}\BibitemShut {NoStop}%
\bibitem [{\citenamefont {Morozov}\ and\ \citenamefont
  {Lin}(2009)}]{morozov09}%
  \BibitemOpen
  \bibfield  {author} {\bibinfo {author} {\bibfnamefont {A.~N.}\ \bibnamefont
  {Morozov}}\ and\ \bibinfo {author} {\bibfnamefont {S.~H.}\ \bibnamefont
  {Lin}},\ }\href {\doibase 10.1063/1.3077658} {\bibfield  {journal} {\bibinfo
  {journal} {J. Chem. Phys.}\ }\textbf {\bibinfo {volume} {130}},\ \bibinfo
  {pages} {074903} (\bibinfo {year} {2009})}\BibitemShut {NoStop}%
\bibitem [{Note1()}]{Note1}%
  \BibitemOpen
  \bibinfo {note} {This scaling formula holds for most potentials when the
  perturbation is defined like this.}\BibitemShut {Stop}%
\bibitem [{\citenamefont {Westergren}\ \emph {et~al.}(2003)\citenamefont
  {Westergren}, \citenamefont {Nordholm},\ and\ \citenamefont
  {Rosen}}]{westergren03}%
  \BibitemOpen
  \bibfield  {author} {\bibinfo {author} {\bibfnamefont {J.}~\bibnamefont
  {Westergren}}, \bibinfo {author} {\bibfnamefont {S.}~\bibnamefont
  {Nordholm}}, \ and\ \bibinfo {author} {\bibfnamefont {A.}~\bibnamefont
  {Rosen}},\ }\href {\doibase {10.1039/B208653K}} {\bibfield  {journal}
  {\bibinfo  {journal} {{Phys. Chem. Chem. Phys.}}\ }\textbf {\bibinfo {volume}
  {{5}}},\ \bibinfo {pages} {136} (\bibinfo {year} {{2003}})}\BibitemShut
  {NoStop}%
\bibitem [{\citenamefont {Lado}(1981)}]{lado81}%
  \BibitemOpen
  \bibfield  {author} {\bibinfo {author} {\bibfnamefont {F.}~\bibnamefont
  {Lado}},\ }\href@noop {} {\bibfield  {journal} {\bibinfo  {journal} {J. Chem.
  Phys.}\ }\textbf {\bibinfo {volume} {75}},\ \bibinfo {pages} {5461} (\bibinfo
  {year} {1981})}\BibitemShut {NoStop}%
\bibitem [{\citenamefont {Ray}\ and\ \citenamefont {Zhang}(1999)}]{ray99}%
  \BibitemOpen
  \bibfield  {author} {\bibinfo {author} {\bibfnamefont {J.~R.}\ \bibnamefont
  {Ray}}\ and\ \bibinfo {author} {\bibfnamefont {H.}~\bibnamefont {Zhang}},\
  }\href {\doibase 10.1103/PhysRevE.59.4781} {\bibfield  {journal} {\bibinfo
  {journal} {Phys. Rev. E}\ }\textbf {\bibinfo {volume} {59}},\ \bibinfo
  {pages} {4781} (\bibinfo {year} {1999})}\BibitemShut {NoStop}%
\bibitem [{\citenamefont {Shirts}\ \emph {et~al.}(2006)\citenamefont {Shirts},
  \citenamefont {Burt},\ and\ \citenamefont {Johnson}}]{shirts06}%
  \BibitemOpen
  \bibfield  {author} {\bibinfo {author} {\bibfnamefont {R.}~\bibnamefont
  {Shirts}}, \bibinfo {author} {\bibfnamefont {S.}~\bibnamefont {Burt}}, \ and\
  \bibinfo {author} {\bibfnamefont {A.}~\bibnamefont {Johnson}},\ }\href@noop
  {} {\bibfield  {journal} {\bibinfo  {journal} {J. Chem. Phys.}\ }\textbf
  {\bibinfo {volume} {125}},\ \bibinfo {pages} {164102} (\bibinfo {year}
  {2006})}\BibitemShut {NoStop}%
\bibitem [{\citenamefont {Sutton}\ and\ \citenamefont {Chen}(1990)}]{sutton90}%
  \BibitemOpen
  \bibfield  {author} {\bibinfo {author} {\bibfnamefont {A.~P.}\ \bibnamefont
  {Sutton}}\ and\ \bibinfo {author} {\bibfnamefont {J.}~\bibnamefont {Chen}},\
  }\href@noop {} {\bibfield  {journal} {\bibinfo  {journal} {Philos. Mag. A}\
  }\textbf {\bibinfo {volume} {50}},\ \bibinfo {pages} {45} (\bibinfo {year}
  {1990})}\BibitemShut {NoStop}%
\bibitem [{\citenamefont {\c{C}agin}\ \emph {et~al.}(1998)\citenamefont
  {\c{C}agin}, \citenamefont {Kimura}, \citenamefont {Qi}, \citenamefont {Li},
  \citenamefont {Ikeda}, \citenamefont {Johnsonb},\ and\ \citenamefont
  {Goddard}}]{cagin99}%
  \BibitemOpen
  \bibfield  {author} {\bibinfo {author} {\bibfnamefont {T.}~\bibnamefont
  {\c{C}agin}}, \bibinfo {author} {\bibfnamefont {Y.}~\bibnamefont {Kimura}},
  \bibinfo {author} {\bibfnamefont {Y.}~\bibnamefont {Qi}}, \bibinfo {author}
  {\bibfnamefont {H.}~\bibnamefont {Li}}, \bibinfo {author} {\bibfnamefont
  {H.}~\bibnamefont {Ikeda}}, \bibinfo {author} {\bibfnamefont {W.~L.}\
  \bibnamefont {Johnsonb}}, \ and\ \bibinfo {author} {\bibfnamefont {W.~A.}\
  \bibnamefont {Goddard}},\ }\href {http://dx.doi.org/10.1557/PROC-554-43}
  {\bibfield  {journal} {\bibinfo  {journal} {MRS Proceedings}\ }\textbf
  {\bibinfo {volume} {554}} (\bibinfo {year} {1998})}\BibitemShut {NoStop}%
\bibitem [{Note2()}]{Note2}%
  \BibitemOpen
  \bibinfo {note} {On such cores, what takes 200 CPU minutes on the author's
  laptop takes around 80. On another laptop with a 2.50 GHz ``Intel Core
  i5-2450'' CPU, this reduces to 45 minutes.}\BibitemShut {Stop}%
\bibitem [{\citenamefont {Paradis}\ \emph {et~al.}(2008)\citenamefont
  {Paradis}, \citenamefont {Ishikawa},\ and\ \citenamefont
  {Koike}}]{paradis08}%
  \BibitemOpen
  \bibfield  {author} {\bibinfo {author} {\bibfnamefont {P.~F.}\ \bibnamefont
  {Paradis}}, \bibinfo {author} {\bibfnamefont {T.}~\bibnamefont {Ishikawa}}, \
  and\ \bibinfo {author} {\bibfnamefont {N.}~\bibnamefont {Koike}},\ }\href
  {http://www.springerlink.com/content/v83170035255514g/} {\bibfield  {journal}
  {\bibinfo  {journal} {Gold Bull.}\ }\textbf {\bibinfo {volume} {41}},\
  \bibinfo {pages} {242} (\bibinfo {year} {2008})}\BibitemShut {NoStop}%
\bibitem [{\citenamefont {Stull}(1947)}]{stull47}%
  \BibitemOpen
  \bibfield  {author} {\bibinfo {author} {\bibfnamefont {D.~R.}\ \bibnamefont
  {Stull}},\ }\href@noop {} {\bibfield  {journal} {\bibinfo  {journal} {Ind.
  Eng. Chem.}\ }\textbf {\bibinfo {volume} {39}},\ \bibinfo {pages} {517}
  (\bibinfo {year} {1947})}\BibitemShut {NoStop}%
\end{thebibliography}


\end{document}